\documentclass[epj]{svjour}
\usepackage{graphics}
\usepackage{psfrag}
\usepackage{times}
\usepackage{amsmath, amssymb}
\usepackage{subfigure,rotating,float}

\def\Sz{\langle S_2 \rangle}
\def\Szp{\langle S^+_2 \rangle}
\def\Szm{\langle S^-_2 \rangle}

\begin{document}
\renewcommand{\arraystretch}{1.2}

\title{A surprising method for polarising antiprotons.} 

\author{Th.~Walcher\inst{1,2}, H. Arenh\"ovel\inst{1}, K. Aulenbacher\inst{1},
  R.~Barday\inst{1} and A. Jankowiak\inst{1}}

\institute{
  Institut f\"ur Kernphysik, Johannes Gutenberg-Universit\"at Mainz,
  D-55099 Mainz, Germany \and
  Laboratori Nazionali di Frascati, Istituto Nazionale di Fisica Nucleare,
  I-00044 Frascati (Rome), Italy}

\date{Received: date / Revised version: date}

\abstract{We propose a method for polarising antiprotons in a storage ring 
  by means of a polarised positron beam moving parallel to the antiprotons. 
  If the relative velocity is adjusted to $v/c \approx 0.002$ the cross 
  section for spin-flip is as large as about $2 \cdot 10^{13}$\,barn as shown
  by new QED-calculations of the triple spin-cross sections. Two possibilities  
  for providing a positron source with sufficient flux density are presented. 
  A polarised positron beam with a polarisation of 0.70 and a flux density 
  of approximately $1.5 \cdot 10^{10}$/(mm$^2$ s) appears to be feasible by 
  means of a radioactive $^{11}$C dc-source. A more involved proposal is the 
  production of polarised positrons by pair production with circularly 
  polarised photons. It yields a polarisation of 0.76 and requires the 
  injection into a small storage ring. Such polariser sources can be used at 
  low (100\,MeV) as well as at high (1\,GeV) energy storage rings providing 
  a time of about one hour for polarisation build-up of about $10^{10}$ 
  antiprotons to a polarisation of about 0.18. A comparison with other 
  proposals show a gain in the figure-of-merit by a factor of about ten.}   

   
\maketitle

\section{Introduction                                            \label{intro}}
The spin of elementary particles is essential for their symmetry and in the
dynamics of their interaction. After better and better experimental methods 
for polarising a broad range of different particles have been developed in 
the last decades, experiments with spin variables represent one of the most
significant methods for investigations in subatomic physics. 

However, the particularly significant case of the antiproton is still not
available for experimental investigations due to the lack of a source of
polarised antiprotons. Such a source would allow for studies with isospin 
and spin symmetry in the interaction of nucleons at low, medium and high
energies. In order to substantiate this, three fields are mentioned:
\begin{enumerate}
\item Spectroscopy of hadrons:\\
The annihilation of antiprotons on protons produces a multitude of final
states with two or more mesons (for a summary see \cite{Klempt:2005pp}).
They carry the potential of containing new states, so called exotics, like 
glue balls or hybrid states composed of quarks and gluons. However, the
analysis of the final states is hampered by the need to perform a partial wave
analysis which is frequently not unique. The exploitation of spin degrees
of freedom for both the projectile antinucleon and the target nucleon would at
least halve the contributing amplitudes and increase the significance of the
search for exotics considerably.

Furthermore, the study of known states would be more selective
making an identification clearer and offer an additional parameter in the
decay dynamics.      
\item Antinucleon-nucleon scattering and reactions:\\
The same arguments hold for antinucleon-nucleon elastic scattering.
These cross sections have been measured from close to threshold up to many
GeV. One particularly intriguing aspect is the spin and isospin dependence of
the antinucleon-nucleon interaction at low energies. As is well known the 
nucleon-nucleon and antinucleon-nucleon potentials are connected in the 
still very successful meson exchange description by the G-parity symmetry. 
However, whereas it appears that the long range part of the potential is in
this way reasonably well described by pion exchange, there is no
sensitivity to the short range part attributed to the vector meson exchange,   
since the annihilation dominates for radii shorter than about 0.8~fm 
(for summaries see \cite{Wal88,Klempt:2002ap}). The different spin 
orientations in the entrance channel close to threshold,
where s- and p-wave scattering dominate, will provide sensitivity to 
vector mesons, i.e. to the short range of the real part of the
antinucleon-nucleon interaction. 

For the annihilation dynamics the question whether the quark reorientation or
the gluonic quark fusion-creation mechanism (OZI rule violation) prevails
is not satisfactorily answered. New data with spin degrees of freedom would 
provide very significant constraints.     
\item Antinucleon-nucleon interactions at the parton level:\\
The generalised parton distributions received recently great attention. It
appears that the transversity distribution would become accessible in
reactions of polarised antiprotons and protons (for a summary see e.g. 
\cite{Barone:2001sp}). Of course, the experimental efforts for such a study
would be considerable beyond the realisation of a polarised antiproton source. 
Recently the $\cal{PAX}$ collaboration has proposed just such an investigation 
for the FAIR facility being prepared at GSI, however, without showing an
effective method of polarising antiprotons \cite{Len05}.       
\end{enumerate} 

Though antiprotons are stable all proposals to polarise
them have been less than satisfactory so far. Of the many proposals
\cite{Kri86} only two possibilities have been considered as possibly feasible.   
In the first, the Filtex collaboration has considered the different
attenuation of the spin components of an initially unpolarised antiproton beam  
due to the difference of the singlet and triplet scattering cross sections 
in the interaction with a polarised hydrogen gas target. In a pilot
experiment with protons instead of antiprotons at the Test Storage Ring TSR  
of the Max-Planck-Institute for Nuclear Physics at Heidelberg in 1992 
\cite{Rat93,Zap95,Zap96} this collaboration did indeed find a small effect. 
The rate of polarisation was, however, only $dP_b/dt = (0.0124 \pm 0.0006)/h$, 
a factor of two smaller than expected theoretically. This is interpreted by
Meyer \cite{Mey94} as due to the contribution of the polarised electrons 
in the polarised hydrogen target. Beside the slow polarisation build-up the 
scheme is questionable for antiprotons since the total annihilation cross 
section is two times larger than the elastic scattering and most likely little
spin dependent meaning that the beam will be quickly reduced in intensity. 
Additionally, the Coulomb scattering in the hydrogen gas target reduces the 
beam lifetime greatly. 

The second possibility is the spin transfer in the scattering of initially
unpolarised nucleons/antinucleons from polarised electrons of a gas target 
placed in the coasting beam of a storage ring \cite{Hor94}. This idea has been 
taken up recently \cite{Rat05} and used as the basis for the proposal of the 
$\cal{PAX}$ collaboration \cite{Len05}. However, the internal polarised
hydrogen gas target will Coulomb scatter the antiproton beam and makes the 
use of a very large aperture storage ring mandatory. Therefore, it is not easy 
to characterise the performance of this method with a few numbers. 
Table~\ref{tab:Rat05} tries to transform the presentation of ref.~\cite{Rat05} 
into numbers which can be compared later to the method proposed in this
article.
 
\begin{table}[h]
\centering  
\begin{tabular}{|l||c|c|c|}
\hline 
example & IT$_{a}$ & IT$_{b}$ & IT$_{c}$ \\
\hline \hline
$\Psi_{acc}$/mrad & 50  &  30  & 10 \\
\hline
$\varepsilon$/mm\,mrad & 500 & 180  & 20\\
\hline
T/MeV & 39 & 61 & 167 \\
\hline
p/(MeV/c) & 273 & 344 & 584 \\
\hline
$\beta = v/c$ & 0.28 & 0.34 & 0.53 \\
\hline
$N_{max}^{\Delta Q}$ & $2.7 \cdot 10^{12}$ & $1.5 \cdot 10^{12}$ & $0.6 \cdot 10^{12}$ \\ 
\hline
$N_{max}^R$ & $3.6 \cdot 10^{10}$ & $3.6 \cdot 10^{10}$ & $3.6 \cdot 10^{10}$ \\
\hline
$\tau_{pol}^{50\%}$/hours & 39.8 & 13.9 & 6.3 \\
\hline
$\tau_{AP}$/hours & 16.7 & 4.6 & 1.2 \\
\hline
$N_{pol}^{50\%}$ & $3.3 \cdot 10^9$ & $1.8 \cdot 10^9$ & $1.9 \cdot 10^8$ \\ 
\hline
\end{tabular}
\caption{Parameters for the polarisation with an internal polarised hydrogen
  target method (IT) according to ref.~\cite{Rat05}. $\Psi_{acc}$ is the
  acceptance angle of accelerator needed to accept the Coulomb scattered
  antiprotons, $\varepsilon = \Psi_{acc}^2 \cdot \beta_{target}$ is the 
  emittance (it is here defined without $\pi$), 
  T is the kinetic energy of the beam, p is the momentum of the beam.
  $N_{max}^{\Delta Q}$ is the maximal number of antiprotons allowed by the
  limit for the incoherent tune shift/spread $\Delta Q = 0.015$, $N_{max}^R$ 
  is the maximal number of antiprotons injected in one hour determined by 
  the production rate of antiprotons of $R=10^7 \bar{p}$/s.  
  $\tau_{pol}^{50\%}$ is the time needed to polarise to 50\%, $\tau_{AP}$
  the time in which the number of antiprotons decays to the fraction 1/e, and  
  $N_{pol}^{50\%}$ the number of polarised antiprotons in the ring after 
  $\tau_{pol}^{50\%}$.}      
\label{tab:Rat05}  
\end{table}
The meaning of the parameters is evident with the exception of the important
single particle space charge limit. It has been calculated using the formula 
for coasting beams of ref.~\cite{Bov70}:  
\begin{equation}
  N_{\text{max}}^{\Delta Q} = 2\pi \varepsilon \beta_{lab}^2 \gamma_{lab}^3
  r_p^{-1} \Delta Q,
\label{spch}
\end{equation}
where $\beta_{lab} = v_{lab}/c, \gamma_{lab} = 1/\sqrt{1-\beta_{lab}^2}$,
$r_p = 1.5$\,am is the ``classical proton radius'' and $\Delta Q$ is the 
incoherent tune shift/spread. For common reference the value 
$\Delta Q=0.015$, a value commonly accepted for the operation of storage 
rings, has been chosen. 

The optimisation of these numbers depends on the way the polarised
antiprotons would be used in a certain experiment. We have chosen the
example of a low energy antiproton polariser ring and the high energy 
experimental storage ring HESR of the FAIR project at GSI \cite{Rat05} 
in order to make a direct comparison possible. However, as we shall discuss 
in section~\ref{sec:app}, the method proposed here can be adapted to many 
different situations in storage rings at low and high antiproton energies. 

As a summary we note that a polariser has to be evaluated by considering\\ 
- the polarisation build-up time,\\
- the degree of polarisation after this time,\\ 
- the number of antiprotons available after this time, and\\
- the phase space of the polarised antiprotons.

The evident remedy for avoiding the problems of the internal polarised target
as Coulomb scattering and the large and expensive accelerator acceptance would 
be the interaction of a pure electron beam or target with a pure antiproton 
target or beam, respectively. This idea, considered since the early phase of 
the Low Energy Antiproton Ring LEAR at CERN, was, however, never thoroughly 
pursued since reliable calculations of the cross sections for the polarisation 
transfer adapted to the situation in a storage ring were missing.
However, as will be shown in the following section~\ref{sec:crossection} the
cross sections for polarisation transfer from the electron to the antiproton 
(like signs of charges) are dramatically smaller than the ones for
positron-antiproton transfer (unlike signs of charges). This means that 
a polarised positron source of sufficient flux density is needed. In 
section~\ref{sec:positron-source} we demonstrate that several options exist, 
all meeting or exceeding the required intensity.    
 
The basis of our proposal is a new QED calculation including Coulomb 
distortions presented in section~\ref{sec:crossection} for systems with 
like and unlike charge signs \cite{Are07,Are07a}. It shows that the total 
polarisation-transfer-cross sections at small energies become very large. 
However, as will be discussed in some detail in section~\ref{sec:ring} only
the spin-flip part of the cross section can polarise a beam in a storage 
ring. Therefore the considerations of Horowitz and Meyer \cite{Hor94,Mey94} 
are  not correct since they treat the non-spin-flip part only. This was 
recently correctly pointed out by Milstein and Strak\-hov\-enko 
\cite{Milstein:2005bx}. However, as will emerge in
section~\ref{sec:crossintegr} the statement of these authors that the
spin-flip cross section due to the hyperfine interaction is generally
negligible is not correct. At very small energies this cross section becomes  
very large if one includes the distortions due to the Coulomb attraction of 
particles with unlike electric charges. 

In principle the polarisation transfer cross sections from electrons to 
nucleons have been calculated in the framework of QED since a long time
\cite{Sco59,Sco66,Dom69,Arn81,Arenhovel:1988qh} and extensively been 
used for the measurement of the electric form factor of the
neutron and proton (see e.g. \cite{Glazier:2004ny,Perdrisat:2003xj}). 
However, these calculations are correctly neglecting the spin-flip part, 
being small for the high energiy electrons used for these experiments, 
and consider the non-spin-flip part in the polarisation transfer only.

After the new calculations of the cross sections have been summarised in 
section~\ref{sec:crossection} we present in section \ref{sec:positron-source} 
a discussion of sources of polarised positrons mostly based on a modest 
extrapolation of existing technologies. They will be shown to suffice for a 
realistic scheme in storage rings in section~\ref{sec:ring}. 
In section~\ref{sec:conc} we discuss the figure-of-merit for the different 
proposals and design examples.

\section{Predictions of polarisation transfer cross sections
\label{sec:crossection}}
\newcommand{\beq}{\begin{equation}}
\newcommand{\eeq}{\end{equation}}
\newcommand{\beqa}{\begin{eqnarray}}
\newcommand{\eeqa}{\end{eqnarray}}

In this section we will briefly review the main results of a recent
calculation of the general polarisation transfer cross section for the
scattering of a polarised hadron (proton or antiproton) on a polarised
lepton (electron or positron) at low energies with inclusion of Coulomb
effects \cite{Are07,Are07a}.  

The general differential cross section for electromagnetic hadron-lepton 
scattering with initially polarised hadron and lepton into a state of 
definite final hadron polarisation (all along incoming momentum as 
$z$-axis) without polarisation analysis of the final lepton is given in 
the c.m.\ system by 
\beqa
\frac{d\sigma^h_{\lambda^f_h,\lambda^i_h,\lambda^i_l}(\theta,\phi)}{d\Omega}&=&
\frac{M_l^2M_h^2}{8\pi^2W^2}\nonumber\\
&&\hspace*{-1.5cm}\times \mathrm{Tr}\Big(T^\dagger(\theta,\phi)
\rho^h_f(\lambda^f_h)T(\theta,\phi)\rho^h_i(\lambda^i_h)
\rho^l_i(\lambda^i_l)\Big)\,,\label{xsection}
\eeqa
where the trace refers to the hadron and lepton spin degrees of freedom. The
invariant energy of the hadron-lepton system is denoted by $W=E_h+E_l$ and the
masses of hadron and lepton by $M_h$ and $M_l$, respectively. 
The nonrelativistic density matrices for the initial and final states of
definite spin along the $z$-axis have the form
\beqa
\rho^{h/l}_i(\lambda^i_{h/l})&=&\frac{1}{2}(1+\lambda^i_{h/l}
\sigma_z^{h/l})\,,\\
\rho^{h}_f(\lambda^f_{h/l})&=&\frac{1}{2}(1+\lambda^f_{h/l}
\sigma_z^{h/l})\,, \eeqa
and $\lambda_{h/l}^{i/f}=\pm 1$. For scattering into a state of definite lepton
polarisation without polarisation analysis of the final hadron one
has to replace $\rho^h_f(\lambda^f_h)$ by $\rho^l_f(\lambda^f_l)$. 

For the $T$-matrix we include the Coulomb charge and the spin-dependent
hyperfine interactions. It has the general form
\beq
T=4\pi \alpha Z_l Z_h\Big(\frac{a_c}{{\vec q}^2}-d\,\vec{\sigma}^h\cdot
\vec{\sigma}^l-\vec{\sigma}^h\cdot\stackrel{\leftrightarrow}{D}(\theta,\phi)
\cdot\vec{\sigma}^l\Big)\,,\label{t-matrix}
\eeq
where $\alpha$ denotes the fine structure constant, $\vec q$ the
three-momentum transfer, and $Z_{h/l}$ the hadron and lepton charge,
respectively. Furthermore, the parameters $a_c$, $d$, and the tensor 
$\stackrel{\leftrightarrow}{D}$ depend on what kind of approximation is 
used, i.e. 
\begin{enumerate}
\item[(i)]
Plane wave (PW), corresponding to one-photon-exchange:
\beqa
a_c^{PW}&=&1\,,\\
d^{PW}&=&\frac{2}{3}c^{SS}\,,\\
D_{ij}^{PW}&=&c^{SS}(\widehat q_i\widehat q_j -\frac{1}{3} \delta_{ij})\,,
\eeqa
where $c^{SS}={\mu_h}/({4M_lM_h})$ with $\mu_h$ for the hadron anomalous 
magnetic moment, and $\widehat {\vec q}$ denotes the unit vector along $\vec
q$. 
\item[(ii)]
Distorted wave approximation for the hyperfine contribution (DW):
\beqa
a_c^{DW}&=& {e^{i\phi_c}}\,,\\
d^{DW}&=&N(\eta_c)^2d^{PW}\,,\\
D^{DW}_{ij}&=&\frac{c^{SS}}{4\pi}\int \frac{d^3r}{r^3}
 \psi^{C(-)}_{\vec p^{\,\prime}}(\vec r)^*\nonumber\\
&&(3\hat{r}_i\,\hat{r}_j-\delta_{ij})\psi^{C(+)}_{\vec p}(\vec
r)\label{dij} \,,\label{t_dwba}
\eeqa
where 
\beqa
\phi_c(\theta)&=&-\eta_c \ln(\sin^2(\theta/2))+2\sigma_c\,,\\
\sigma_c&=&\arg(\Gamma(1+i\eta_c))\,,
\eeqa
denotes the Coulomb phase with 
\beq
\eta_c=-\alpha Z_l Z_h/v
\eeq 
as Sommerfeld Coulomb parameter, $v$ as the relative hadron-electron
velocity, and with the normalisation factor 
\beq
N(\eta_c)=\sqrt{\frac{2\pi\eta_c}{e^{2\pi\eta_c}-1}}\,e^{i\sigma_c}\,.
\eeq 
of the incoming and outgoing Coulomb scattering wave functions \cite{Mes69}  
$\psi^{C(\pm)}_{\vec p}$, respectively. 
\end{enumerate}
One should note that the tensor $D_{ij}$ is symmetric and traceless.

Evaluation of the trace in eq.\,(\ref{xsection}) results in the following 
expression for the cross section \cite{Are07a}
\beqa
\frac{d\sigma^h_{\lambda^f_h,\lambda^i_h,\lambda^i_l}(\theta)}{d\Omega}&=&
(1+\lambda_h^i\lambda_h^f)S_0(\theta) \nonumber\\
&&\hspace*{-1.5cm}+\lambda_l^i(\lambda_h^-S_2^-(\theta)
+\lambda_h^+S_{2}^+(\theta))
+\lambda_h^i\lambda_h^fS_2(\theta)\,,
\label{hadron1}
\eeqa
independent of the azimuthal angle $\phi$, where
$\lambda_h^\pm=\lambda_h^i\pm\lambda_h^f$ and 
\beqa
S_0(\theta)&=&V\Big(\frac{|a_c|^2}{{\vec q}^4}
+3|d|^2+|D^0_{11}(\theta)|^2+|D^0_{22}(\theta)|^2
\nonumber\\
&&+|D^0_{33}(\theta)|^2+2|D^0_{13}(\theta)|^2\Big) \,,\label{s0}\\
S_2^+(\theta)&=&2\frac{V}{{\vec q}^2}\Re e(a_c^*(d+D^0_{33}(\theta)))\,,\\
S_2^-(\theta)&=&2V\Big(\Re e(d^*D^0_{33}(\theta)-D^0_{11}(\theta)^*
D^0_{22}(\theta))-|d|^2\Big)\,, \qquad\\
S_2(\theta)&=&2V\Big(2\Re e(d^*D^0_{33}(\theta))-2|d|^2-|D^0_{11}(\theta)|^2
\nonumber\\
&&-|D^0_{22}(\theta)|^2-|D^0_{13}(\theta)|^2\Big)\label{s2}\,,
\eeqa
with
\beqa
V&=&\frac{2\alpha^2 Z_l^2 Z_h^2 M_l^2 M_h^2}{W^2}\,.
\eeqa
The unpolarised cross section is obtained by summing over all spin projections
and division by four, i.e.\
\beq
\frac{d\sigma_0}{d\Omega_h}=2S_0\,.
\eeq

Now we will discuss different aspects of the general expression of
eq.\,(\ref{hadron1}) in order to analyse and distinguish spin-flip and non-spin-flip
contributions to the polarisation transfer. To this end we will consider
completely polarised leptons along the $z$-axis, i.e. $\lambda_l^i=\pm 1$. The
non-spin-flip contributions are given by $\lambda_h^i=\lambda_h^f=\pm 1$
\beqa
\frac{d\sigma^h_{+,+,\lambda^i_l}(\theta)}{d\Omega}&=&
2(S_0(\theta)+\lambda_l^iS_{2}^+(\theta))+S_2(\theta)\,,\label{nospinflip+}\\
\frac{d\sigma^h_{-,-,\lambda^i_l}(\theta)}{d\Omega}&=&
2(S_0(\theta)-\lambda_l^iS_{2}^+(\theta))+S_2(\theta)\,.\label{nospinflip-}
\eeqa
Thus in this case, the polarisation transfer is not the result of a hadron
spin-flip, but arises from different scattering strength for the two spin
orientations determined by the hyperfine interaction. Their difference for
$\lambda_l^i= 1$ is just the quantity
\beq
P_{zz}\frac{d\sigma_0}{d\Omega_h}=
\frac{d\sigma^h_{+,+,+}(\theta)}{d\Omega}
-\frac{d\sigma^h_{-,-,+}(\theta)}{d\Omega}=2S_{2}^+(\theta)\,,
\eeq
which has been considered by Horowitz and Meyer \cite{Hor94}. Although
yielding a polarising effect in the differential cross section, it cannot
produce a net hadron polarisation for the situation when all scattered
particles are collected together with the incoming beam as will be discussed 
in detail in section~\ref{sec:build-up-dc}. This is different for
the genuine spin-flip contribution, i.e.\ $\lambda_h^i=-\lambda_h^f=\pm 1$,
which are given by
\beqa
\frac{d\sigma^h_{-,+,\lambda^i_l}(\theta)}{d\Omega}&=&
2\lambda_l^iS_{2}^-(\theta)-S_2(\theta)\,,\label{spinflip+}\\
\frac{d\sigma^h_{+,-,\lambda^i_l}(\theta)}{d\Omega}&=&
-2\lambda_l^iS_{2}^-(\theta)-S_2(\theta)\,.\label{spinflip-}
\eeqa
We would like to mention that $d\sigma^h_{-,+,\lambda^i_l=+}/{d\Omega}$ had
to vanish if the total spin projection $s_z^h+s_z^l$ would be conserved as
intuition might suggest. But this is not the case because of the tensor part
of the hyperfine interaction in eq.\,(\ref{t-matrix}). Indeed, inserting the 
explicit expressions for $S_{2}^-$ and $S_2$, one finds
\beqa
\frac{d\sigma^h_{-,+,+}(\theta)}{d\Omega}&=&
2V\Big(|D^0_{11}(\theta)-D^0_{22}(\theta)|^2 +|D^0_{13}(\theta)|^2\Big)\,,\quad
\eeqa
which vanishes completely for $D^0_{ij}=0$. For the other spin-flip
contribution one finds
\beqa
\frac{d\sigma^h_{-,+,-}(\theta)}{d\Omega}&=&
2V\Big(|2d-D^0_{33}(\theta)|^2 +|D^0_{13}(\theta)|^2\Big)\,,
\eeqa
which contributes even for vanishing tensor interaction.

It is just the difference of these two spin-flip contributions which leads to
a non-zero net hadron polarisation in a storage ring as will be discussed in
Section 4.1.

\subsection{Results for the integrated spin-flip cross sections}
\label{sec:crossintegr}
Since the relevant quantities for the polarisation build-up in a storage ring
are the integrated spin-flip cross sections we have integrated the general
polarisation transfer cross section of eq.\,(\ref{hadron1}) 
over the solid angle except for the small region $\theta<\theta_{min}$.
The minimal scattering angle is determined by the requirement that the
impact parameter should not exceed a given value b, i.e.\
\beq
\theta_{min}=2\arctan({\eta_c}/{l})\,,
\eeq
with the classical angular momentum $l=pb$. In detail we define
\beqa
\sigma^h_{\lambda^f_h,\lambda^i_h,\lambda^i_l}&=&2\pi\int_{\theta_{min}}^\pi 
d\cos(\theta)\frac{d\sigma^h_{\lambda^f_h,\lambda^i_h,\lambda^i_l}(\theta)}
{d\cos(\theta)}\nonumber\\
&=&(1+\lambda_h^i\lambda_h^f)\langle S_0\rangle\\
&&+\lambda_l^i(\lambda_h^-\langle S_2^-\rangle
+\lambda_h^+\langle S_{2}^+\rangle)
+\lambda_h^i\lambda_h^f\langle S_2\rangle\,.\nonumber\label{inthadron1}
\eeqa
The equations corresponding to eqs.\,(\ref{nospinflip+}),(\ref{nospinflip-}) 
and eqs.\,(\ref{spinflip+}),(\ref{spinflip-}) are

\noindent
(i) no hadron spin-flip
\beqa
\sigma^h_{+,+,\lambda^i_l}&=&
2(\langle S_0\rangle+\lambda_l^i\langle S_{2}^+\rangle)+\langle S_2\rangle
\,,\label{intnospinflip+}\\
\sigma^h_{-,-,\lambda^i_l}&=&
2(\langle S_0\rangle-\lambda_l^i\langle S_{2}^+\rangle)+\langle S_2\rangle
\,.\label{intnospinflip-}
\eeqa
(ii) hadron spin-flip
\beqa
\sigma^h_{-,+,\lambda^i_l}&=&
2\lambda_l^i\langle S_{2}^-\rangle-\langle S_2\rangle\,,
\label{intspinflip+}\\
\sigma^h_{+,-,\lambda^i_l}&=&
-2\lambda_l^i\langle S_{2}^-\rangle-\langle S_2\rangle\,.
\label{intspinflip-}
\eeqa

\begin{figure}[ht]
\includegraphics[width=1.\columnwidth]{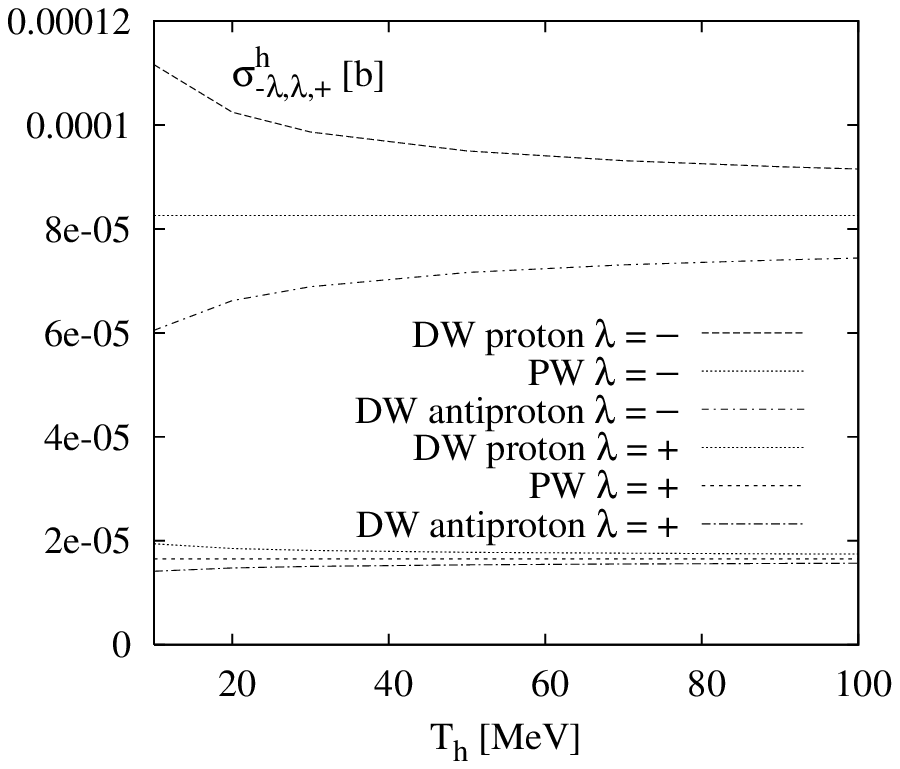}
\caption{The integrated spin-flip cross sections 
$\langle \sigma^h_{-\lambda,\lambda,+}\rangle$ ($\lambda=\pm$) for antiproton
and proton electron scattering in the c.m.\ frame as function of the proton
lab kinetic energy $T_h$ for $b=10^{10}$~fm in PW and DW.}
\label{sigspin-flip100} 
\end{figure}

\begin{figure}[ht]
\includegraphics[width=1.\columnwidth]{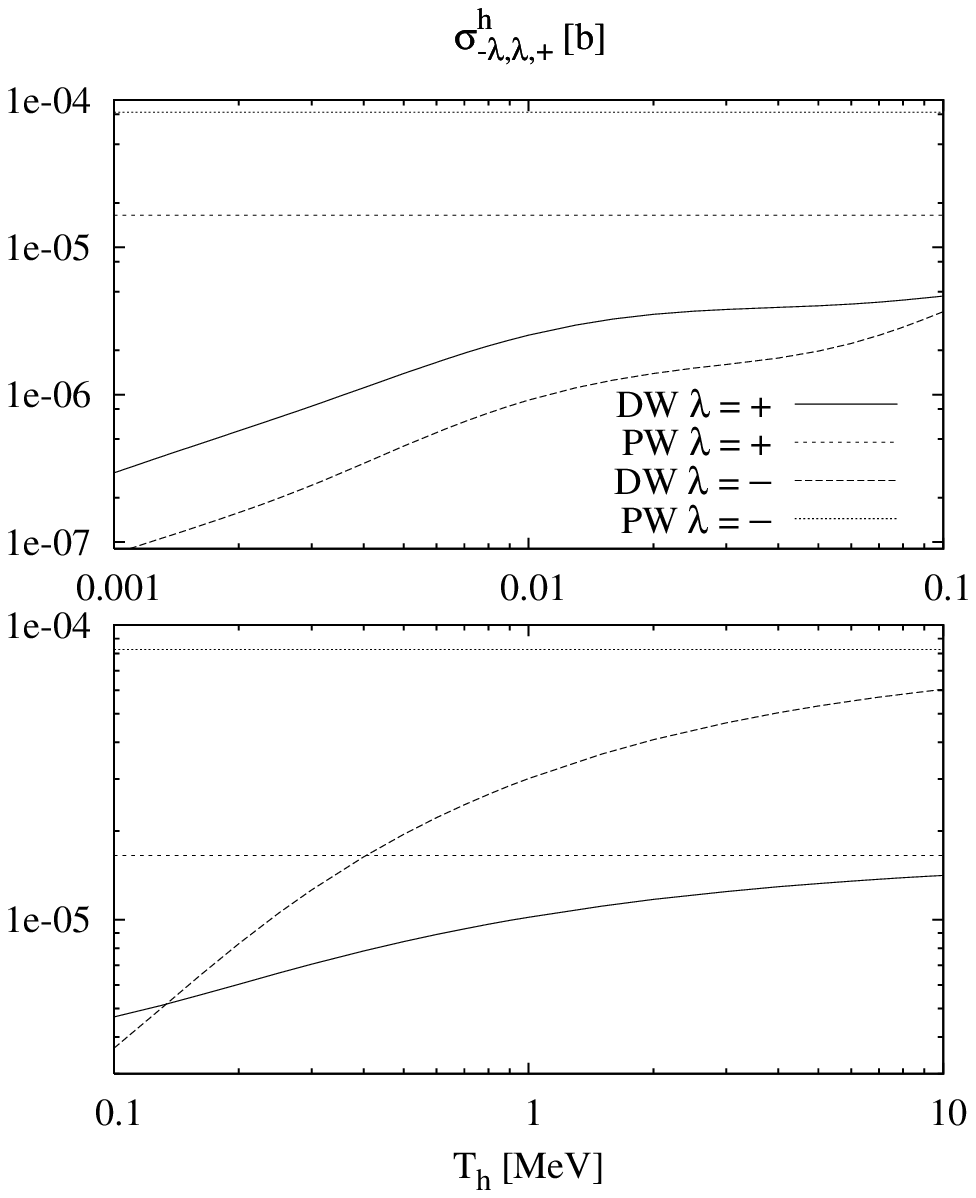}
\caption{The integrated spin-flip cross sections 
$\langle \sigma^h_{-\lambda,\lambda,+}\rangle$ ($\lambda=\pm$) for antiproton
  electron scattering in the c.m.\ frame as function of the proton lab kinetic
  energy $T_h$ for $b=10^{10}$~fm in PW and DW.}
\label{sigspin-flip-} 
\end{figure}

\begin{figure}[ht]
\includegraphics[width=1.\columnwidth]{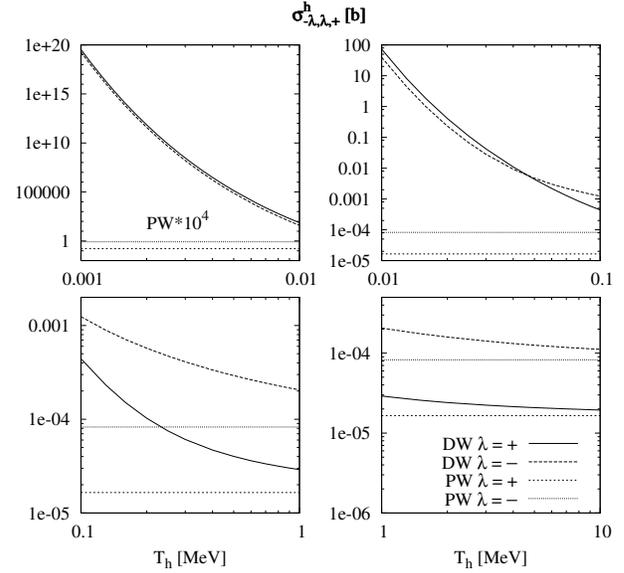}
\caption{The integrated spin-flip cross sections 
$\langle \sigma^h_{-\lambda,\lambda,+}\rangle$ ($\lambda=\pm$) for proton
  electron scattering in the c.m.\ frame as function of the proton lab kinetic
  energy $T_h$ for $b=10^{10}$~fm in PW and DW.}
\label{sigspin-flip+} 
\end{figure}

In Figure~\ref{sigspin-flip100} we show first for the range of higher lab
kinetic  energies between 10 and 100~MeV the result for the two integrated
spin-flip cross sections $\langle \sigma^h_{-\lambda,\lambda,+}\rangle$ 
($\lambda=\pm$) for both, antiproton and proton electron scattering for 
$b=10^{10}$~fm without, i.e.\ in PW, and with inclusion of Coulomb effects  
in the distorted wave approximation (DW). One readily notes, that Coulomb 
effects lead for the proton to an enhancement of the cross sections compared 
to the PW result due to the attraction of the Coulomb field whereas for the 
antiproton the repulsion reduces the cross sections. Furthermore, one notes 
that the Coulomb influence decreases with increasing kinetic energy, as is 
to be expected. 

The lower energy range between 0.001 and 10~MeV is displayed in
Fig.~\ref{sigspin-flip-} for antiproton electron scattering in PW and DW. It
is  apparent that below a kinetic energy of 10~MeV Coulomb effects continue
to strongly suppress both cross sections, the reduction rapidly increasing
with decreasing kinetic energy. 

The corresponding results for the proton case are shown in
Fig.~\ref{sigspin-flip+}. In contrast to the antiproton case one notes here a
very rapid increase of the integrated polarisation cross sections with
decreasing energy. This rapid increase is caused essentially by the
strong attraction of the Coulomb field pulling in the scattering wave
towards small distances. It is governed by a
factor $e^{-2\pi\eta_c}$ (for details see Ref.~\cite{Are07}) which
grows very fast with decreasing energy (i.e.\ increasing $\eta_c$)
because for proton electron scattering one has $\eta_c<0$. 

\begin{center}
\begin{figure}[ht]
\centerline{
\includegraphics[width=0.8\columnwidth]{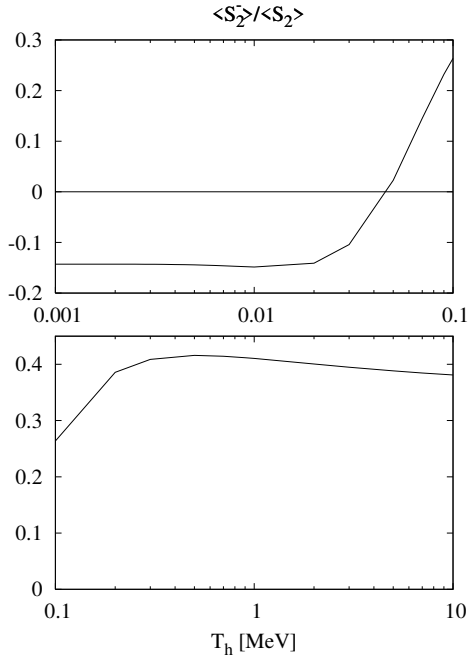}}
\caption{Ratio of $\langle S_2^-\rangle/\langle S_2\rangle$ for proton
  electron scattering in the c.m.\ frame as function of the proton lab kinetic
  energy $T_h$ for DW approximation.}
\label{ratiospin-flip+} 
\end{figure}
\end{center}
Finally, we show in Fig.~\ref{ratiospin-flip+} the ratio $\langle
S_2^-\rangle/\langle S_2\rangle$ for proton electron scattering which governs
the final proton net polarisation (see Section 4).

\section{Source of polarised positrons}                       
\label{sec:positron-source}
\subsection{Direct current (dc) positron source}
\label{sec:positron-source-dc}
DC polarised positron beams of narrow energy spread ($\delta E<
2$\,eV)  emitted from small source areas ($A_{\text{source}}=0.1$\,cm$^2$)   
have been developed more than 20 years ago at the University of
Michigan \cite{VanHouse}. The working principle is angular
selective absorption in low Z absorbers with subsequent moderation
in high-purity metal sheets.

The Michigan source  was based on a $^{22}$Na emitter ($\beta^+$
endpoint energy 0.544\,MeV) with an activity of 1.7\,GBq. The
absorption/moderation process limited the beam intensity to
$5 \cdot 10^5 e^+$/s with a polarisation of $P_{e^+}=0.48$. The moderator 
efficiency depends on the absence of positron trapping in defect
sites in the bulk of the moderator. 

Moderator efficiencies which are larger by more than two orders of
magnitude  have been obtained by using laser annealed thin
tungsten foils in ultra high vacuum \cite{Jac90}. Furthermore, the
source activity can be increased by on-line radio-isotope
production with a dedicated ion accelerator.

A suitable candidate isotope is  $^{11}$C  which is produced by
the $^{14}$N(p,$\alpha)^{11}$C reaction. Typical saturation yields
from a Nitrogen gas target are 8\,GBq/$\mu$A at a proton energy 
of 18\,MeV \cite{VDC85}. Proton linear accelerators achieving 
more than 1\,mA in the desired energy range are commercially 
available \cite{NAG06}. Therefore, a source activity of
$10^{13}$\,Bq seems to be achievable at reasonable  investment
and operating costs. This means that the introduction of improved 
moderators and online radioisotope production  will allow an 
increase of the positron intensity by six orders of magnitude 
compared to the Michigan source. 

One can make use of this increased intensity to obtain lower positron 
beam emittance and higher polarisation in the following way:
The activity will be extracted and deposited onto a source area of
about 0.3\,mm diameter. In addition it seems advisable to increase
beam polarisation by stronger selective absorption. Due to the
larger $\beta^+$ endpoint energy of $^{11}$C (1.0\,MeV versus 0.544\,MeV for
the vast majority of the positrons emitted by $^{22}$Na) and the increased 
selective absorption we expect an increase  of beam polarisation towards  
$P_{e^+}=0.7$. We believe that  the losses inflicted  by these two measures
will reduce the overall gain with respect to the Michigan design from $10^6$
to $10^4$  yielding an intensity of $5 \cdot 10^9$e$^+$/s.

An advantageous feature of moderated positrons is their very low
energy spread. Positron beam temperatures of less than 100\,K have 
been observed when the moderator was cooled to 23\,K \cite{FIS86}. 
Since the normalised beam emittance scales $\propto \sqrt{T}$ a much 
smaller emittance per emitting area compared to a thermionic (electron)
cathode can be obtained. The geometric emittance of the polarised
positron beam generated from  the 0.3 mm diameter spot at the surface 
will therefore approach the emittance of the cooled antiproton beam
(see section \ref{sec:app}). This means that additional solenoid focusing  
of the positron beam could then be given up offering complete
freedom for choosing the spin direction in the interaction region of 
positrons and antiprotons. Since the initial energy of the positron beam 
is very low the desired spin orientation of the positrons can be obtained 
with very compact spin manipulators as already demonstrated by the Michigan 
group. Following the spin manipulation an electrostatic post-accelerator 
will provide the kinetic energies which are required for the two design
examples presented in Table~\ref{tab:table_PB}.

\subsection{Pulsed positron source with a storage ring} 
\label{sec:positron-source-pulsed}
A more effective use of polarised positrons can be made by injecting 
them as a pulse into a low energy storage ring. Low energy lepton storage 
rings have already been proposed \cite{LEPTA1,LEPTA2} and are expected 
to work stably for several seconds at currents of many mA. In order to
achieve single turn injection the pulse length should be well below the
revolution time. For our purposes we assume a revolution time of 50\,ns 
corresponding to ~14\,m circumference for stored positrons of 1\,MeV kinetic 
energy. As will be discussed in section~\ref{sec:app} the momentum spread 
of the beams should be $\pm 1 \cdot 10^{-4}$ setting a limit to the 
longitudinal acceptance of $\delta E \delta t < 5$\,keV\,ns.

\begin{figure}[h]
\begin{center}
\includegraphics[width=0.85\columnwidth]{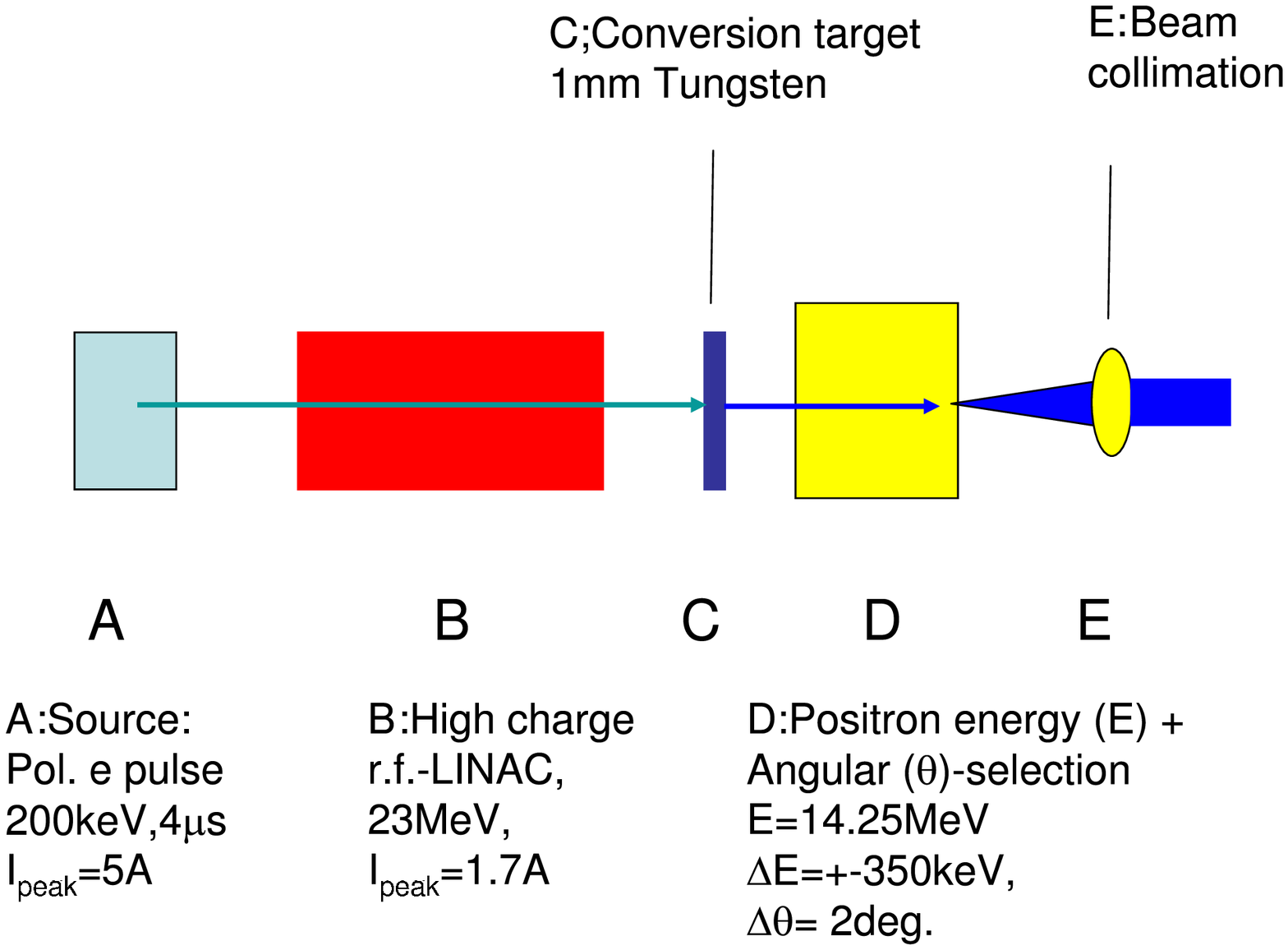}\\[2mm]
\includegraphics[width=0.85\columnwidth]{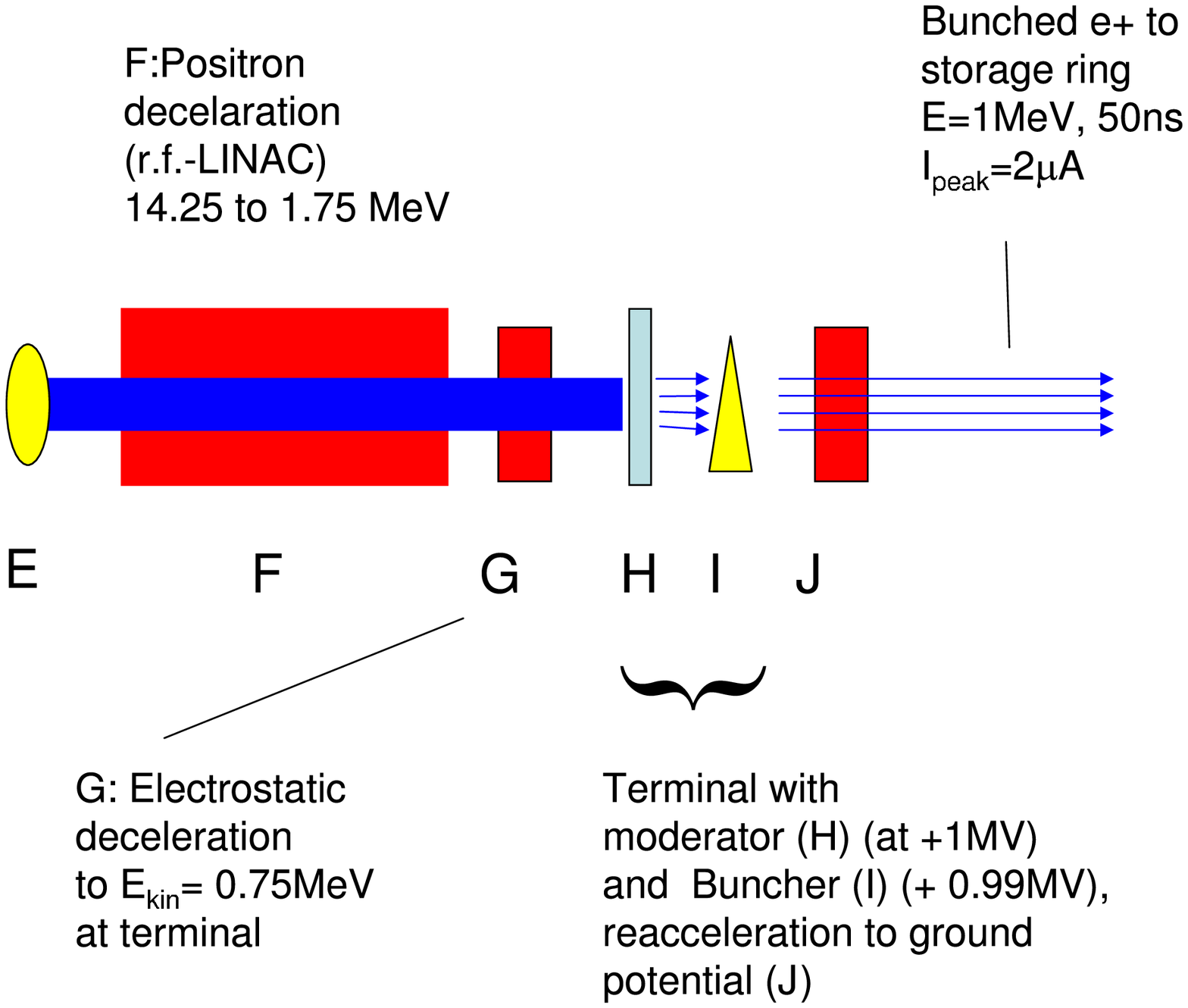}
\end{center}
\caption{The different stages of the pulsed positron source. A: source of
  polarised electrons, B: linear electron rf-accelerator for acceleration to
  23\,MeV, C: combined bremsstrahl and positron production target, D: momentum
  analyser and solid angel selection, E: focusing into the linear positron 
  rf-decelerator, F: linear positron rf-decelerator, G: electrostatic 
  decelerator, H\&I: terminal at +1\,MV high voltage with moderator (H) and
  buncher (I), J: electrostatic accelerator.}
\label{fig:pulse_schema} 
\end{figure}

As is well known longitudinally polarised positrons can be generated by
converting circularly polarised bremsstrahlung in a converter target 
\cite{Olsen:1959}. The polarised bremsstrahlung in turn is produced from
impinging a beam of polarised electrons on a bremsstrahl target. Here we 
propose one scheme working with existing technology and describe it starting
from the source of polarised electrons to the injection into the mentioned 
positron storage ring. In order to make it easier to follow 
Fig.~\ref{fig:pulse_schema} shows the different stages in the sequence of the
following description.  

The polarised electrons are produced by photo-exciting a suitable 
semiconductor heterostructure, e.g. strained GaAs.  Operation of such a
source with current densities of several A/cm$^2$ has already been
demonstrated \cite{SLAC96,SLAC04}. For reasons which will become clear 
shortly we need a $4 \mu$s long driver pulse with 1.7~A peak current 
($6.8\,\mu$C or $4.3 \cdot 10^{13}$ electrons per $4\,\mu$s-pulse). 
The polarisation produced routinely with such sources is $P_{e^-}=0.83$. 
We choose an electron beam energy of 23\,MeV which can be generated by 
a linear rf-accelerator. The mentioned energy and bunch charge parameters 
have already been demonstrated at CERN using optimised rf-travelling wave 
structures of rather compact design \cite{CTF04}. A capture efficiency of 
larger than 0.75 was achieved necessitating a peak current of up to 2.2\,A at  
the source. As will be discussed later we project a 28\,Hz repetition rate 
yielding an average current of $250\,\mu$A of polarised electrons which is 
a factor of two larger than that of the routine operation at MAMI 
\cite{Aul97,Aul04}, well within reach of ongoing development as discussed 
below. The $4\,\mu$s pulse can easily be produced at the source of polarised
electrons by gain switching a commercial high power semiconductor laser. 

After acceleration the electron beam will be focused on a Tungsten
($Z=74, A_r=183.8$) bremsstrahl target of 1\,mm thickness. From the 
Bethe-Heitler cross section we estimate for the given peak current 
of 1.7\,A a flux of $1.2 \cdot 10^{13}$\,photons/($4 \mu$\,s-pulse) 
for the energy interval from 16 to 23\,MeV in a cone with a polar angle 
of $2^{\circ}$. All photons in this interval can produce an electron-positron 
pair with a total energy of the positron of $E_+ = (p_+^2 + m_ec^2)^{1/2} = 
15.5\,$MeV. 

As we shall see in section~\ref{sec:app} we need a good definition of 
the kinetic energy of the positrons. Therefore, we cut out of the broad 
positron spectrum a bite of $\delta\,p/p_+ = \pm 2.5 \cdot 10^{-2}$ or 
$\delta\,E_+ \approx \delta\,p_+ = 700$\,keV. The design of a suitable 
beam-handling system allowing this cut poses no problem.  

We now use the bremsstrahl target also as conversion target of the circularly
polarised photons into longitudinally polarised positrons. The positrons will
be produced over the full length of the target meaning that the effective
target thickness is 0.5\,mm. Due to collision energy losses these positrons  
will be decelerated to 14.25\,MeV before leaving the target. We extrapolate 
the differential cross section for the conversion from Fig.~6.12 in 
ref.~\cite{Motz:1969ti} for $\theta_+ = 0^{\circ}$ in the direction of the 
photon. We multiply with the momentum bite of 700\,keV, the solid angle of 
$7.7 \cdot 10^{-3}$ (opening angle of forward cone $2^{\circ}$), and the 
photon flux, yielding a peak positron flux of 
$8.5 \cdot 10^{10}$\,e$^+ /(4 \mu$s-pulse).

So far we have neglected the multiple Coulomb scattering leading to a 
widening of the positron angular distribution in forward direction.The 
positrons will be produced over the full length of the target, i.e. 
an effective target thickness of 0.5\,mm.  The rms-angle of this 
distribution can be calculated from the standard formula in 
ref.~\cite{Lynch:1991xx} to be $17^{\circ}$. Integrating over the 
$2^{\circ}$ cone gives a correction factor of $6.8 \cdot 10^{-3}$ resulting 
in $3.9 \cdot 10^{7}$\,e$^+ /(4 \mu$s-pulse). 

For $\langle E_{\gamma} \rangle /E_e \approx 18.5/23$ one gets a degree of
polarisation of the photons of 0.95 (see Fig.~5 in ref.~\cite{Olsen:1959}). 
Multiplying this value with the polarisation of the electrons results in 0.79 
for the polarisation of the photons. The polarisation of the positrons can now 
be determined by realising that $\langle E_{+} \rangle /E_{\gamma} \approx
15.5/18.5 = 0.84$. From Fig.~5.05 and 5.06 of ref.~\cite{Motz:1969ti} one
reads off $P^L_+ = 0.96$ so that we end with an effective polarisation of 
the positrons of $P_{e^+}=0.76$. We estimate polarisation losses due to 
large angle Mott-scattering in the positron production target to be
negligible. 

However, after passing the momentum selection the posi\-trons still 
have a energy spread too large to be usable in the positron 
storage ring. At this stage we can make use of the moderation techniques 
already discussed for the dc-source in section~\ref{sec:positron-source-dc}.  
The positrons are decelerated by a further linear accelerator from 15.5\,MeV  
down to about 1.75\,MeV and then further to 750\,keV by means of a high
voltage stage. In the terminal of this high voltage stage we place the moderator. 
After passing the moderator the positrons are accelerated again up to the 
optimal kinetic energy needed for the spin transfer. For our example discussed  
in section~\ref{sec:HESR} this energy is $\approx 1$\,MeV which the positron 
beam acquires by leaving the high voltage stage towards ground potential.

The initial transverse momenta are relatively large, thus beam guidance has to   
be provided in the decelerator by a strong longitudinal solenoid field. For 
the parameters discussed in the following a solenoid field of 0.5\,T suffices.    
Before entering the decelerator the transverse momenta will be reduced by 
expanding the beam and collimating it with a suitable lens. Assuming an 
increase of the beam diameter from initially 2 to 10\,mm this will lead to 
maximum beam angles at the moderator surface of less than $11^{\circ}$.  
The depolarisation originating from the different orientations of the positron
momenta can be neglected. The transverse emittance of the positron beam
leaving the moderator will of course be much larger than in the case of the
dc-source, but this is partially compensated for by the higher beam energy at
the exit. The geometric emittance for the injection into the positron storage
ring will therefore be about $3 \pi$\,mm\,mrad. It is important to note that  
an energy width of less than 0.1\,eV is obtained after the moderation 
resulting in a longitudinal phase space of $\varepsilon_L \approx 
0.4$\,keV\,ns  well below the acceptance of the positron storage ring. 
Before injection the positron pulse must be compressed in time by a 
factor of 80 to 50\,ns by a bunching system which, however, is not 
problematic.

Compared to a radioactive $\beta^+$-source we find that the decelerated 
positron beam when hitting the moderator has a similar energy and energy 
width but is better collimated. We therefore expect that the overall 
efficiencies for moderation of radioactive sources of about 0.7\% 
\cite{Schultz:1988yy} can be doubled. Multiplying the total number of 
highly  polarised positrons given above with the estimated moderator 
efficiency of 0.015 leaves us with $6 \cdot 10^5$\,e$^+ /(4 \mu$s-pulse). 
It is interesting to note that a very similar positron target has been 
proposed in ref.~\cite{Potylitsin:1997zz}. 

We therefore project the injection of pulses with $6 \cdot 10^5$  positrons 
into the low energy storage ring which results in  a stored current of
$1.2 \cdot 10^{13}$e$^+$\,s$^{-1}$ or 2\,$\mu$A, exceeding the value of 
the dc-source by a factor of 2000. As will be discussed in detail in 
the next section the storage ring must be refilled as soon as a 
considerable fraction of positrons has transferred their polarisation to the  
antiprotons. Thus a high repetition rate of the driver pulse is demanded 
for. The limiting factor for the driver rate is the photocathode lifetime.
Presently operating conditions at MAMI allow for the extraction of 200\,C
within one lifetime of the photocathode \cite{Aul04} yielding 220 hours of 
continuous operation for the 250\,$\mu$A average current required here. 
Since photocathode regeneration can be obtained within less than two hours 
by the techniques developed at MAMI \cite{Aul97} it follows that downtime 
will be negligible.
 
We have already performed test experiments with average polarised beam
currents exceeding 10\,mA \cite{Barday06}. These results suggest that
operation of a polarised source with sufficiently long lifetime at even an 
average current of 30\,mA is possible. This corresponds to a repetition 
frequency of 1500\,Hz with the same number of positrons per pulse as above. 
However, one has to optimise the peak current, the repetition frequency, 
i.e. the average current, and the cathode lifetime. Since this depends on 
parameters specific for different applications we have not done this 
optimisation.

If even higher currents of stored polarised positrons are needed 
one can resort to the more expensive concept of radiative polarisation
(Sokolov-Ternov effect) \cite{Derbenev}. For the VEPP-2000 project
(a storage ring of 4\,m radius operating at 1\,GeV) it is
envisaged to polarise $10^{11}$ positrons within 12 minutes by
radiative polarisation \cite{Shatunov}. After polarisation a
considerable fraction of the positrons could be decelerated and 
injected into the low energy storage ring. 

\section{Realisation with a storage ring                       
\label{sec:ring}}
After the two essential ingredients, the polarisation-transfer \\ cross
sections and the source of polarised positrons have been presented, we want 
to show how one can build an antiproton polariser based on them. The basic
idea is to let the two beams move in parallel with a small relative velocity,  
i.e. a small relative kinetic energy, and take advantage of the large 
polarisation-transfer cross section of the leptons to the hadrons in forward 
direction, making up for the relatively low density of the beam of polarised 
positrons. (In the following we follow the notation of 
section~\ref{sec:crossection}. Therefore the lepton stands the for
positron/electron and the hadron for the antiproton/proton, respectively.)
\subsection{Polarisation build-up}                           
\label{sec:build-up}
\subsubsection{Polarisation build-up with a dc-positron source.} 
\label{sec:build-up-dc}
Firstly, we want to calculate the polarisation build-up in an antiproton
storage ring using a dc-positron source, the beam of which overlaps with the 
antiproton beam. We go into the rest frame of the circulating antiproton beam 
and can use the results of section~\ref{sec:crossection} directly since the
cross sections transform as scalars. The quantities in this system are
indicated by primes.

Though it is not mandatory for all cases (see section~\ref{sec:positron-source}) 
we assume to transport the polarised lepton beam by means of a solenoid with
its field in the longitudinal beam direction in order to guarantee a diameter 
of 2\,mm over the full overlap length and to guide the positron beam. 
Consequently, we chose the quantisation axis (longitudinal spin direction) in 
the direction of the solenoid field. This means one has to provide a Siberian 
snake in the ring for rotating the spin longitudinally at the entrance of the 
solenoid after one circulation. 

For the application of eqs.\,(\ref{intspinflip+}) and (\ref{intspinflip-}) we 
have firstly to recall the definition of the polarisation of a particle beam:
\begin{equation}
P_{beam}= \frac{N_{+}-N_{-}}{N_{+}+N_{-}} 
\end{equation}
where $N_{+/-}$ denotes the number of particles with spin parallel and
antiparallel to the polarisation axis, respectively. This definition means 
that we have to find a difference of the cross sections for the two 
spin projections along the polarisation axis in a certain kinematical 
situation. We realise easily that the interactions of case (i) (no spin-flip) 
cannot polarise the hadron beam in a storage ring. The maximal scattering 
angle of antiprotons from an electron target is $\theta_{max}' = m_e/m_p 
\approx 0.5\,$mr. This angle has to be transformed into the ring system 
giving $\theta_{max} \ll \theta_{acc}$ where $\theta_{acc}$ is the acceptance 
angle of the storage ring. This means that due to the kinematics in the 
storage ring all particles, whether scattered or not, are recollected in 
the beam and, therefore, the polarisation of the coasting beam is unchanged. 
Another way of looking at this situation is to imagine that the scattered 
hadrons are ``taken out of the beam'' \cite{Nikolaev:2006gw}. The scattered 
hadron ensemble is then indeed  polarised due to the different cross sections 
for opposite hadron spin projections under a finite scattering angle in a 
certain solid angle (see eq.\,(\ref{hadron1}). These hadrons are missing in 
the hadron beam and leave the beam with a polarisation just, equal but
opposite in sign, to the polarisation of the scattered hadrons. Since the scattered
hadrons and the remaining beam are not separated geometrically but stay in 
the coasting beam, no polarisation can be build up on the basis of the cross 
sections of case (i) in subsection~\ref{sec:crossintegr}.

On the other hand, the cross sections of case (ii) in 
subsection~\ref{sec:crossintegr} are describing a spin-flip, 
i.e. the hadrons interacting according to these change the spin projection. 
However, only if the spin-flip is asymmetric for a given lepton polarisation 
$\lambda^l_i$ a net polarisation within the coasting beam can be achieved. 
This is, however, just the case here since 
\begin{equation}
\sigma^h_{-,+, \lambda^l_i} \neq \sigma^h_{+,-, \lambda^l_i}.
\end{equation}

The distinction of case (i) and (ii) in section \ref{sec:crossintegr} 
has been obscured in the past by an unfortunate nomenclature. The polarisation 
transfer, due to the scattering without spin-flip has been called ``spin
transfer'' \cite{Hor94,Mey94} suggesting spin-flip. The fact that the term 
$\Szp$ is not describing spin-flip has been pointed out by Milstein and 
Strakhovenko \cite{Milstein:2005bx} as already mentioned in the introduction. 
In an earlier version the authors of this paper \cite{Walcher:2007sj}
following  refs.\,\cite{Hor94,Mey94} made the same mistake and used their 
$\langle P_{zz} \sigma \rangle$ which is the same as the 
$d \sigma/d \Omega K_{j00i}$ of ref.\,\cite{Hor94}. Therefore, as shown in 
section \ref{sec:crossection}, it is not enough to consider the polarisation 
transfer, i.e. cross sections with two spin degrees of freedom, but one has 
to study the triple-spin-cross sections. 

We now can proceed and calculate the rate of change of the number of particles
with a given hadron spin projection in the ring. First, we consider all
quantities in the rest frame of the leptons and indicate them by a prime. We 
apply the eqs.\,(\ref{intspinflip+}) and (\ref{intspinflip-}) to the macroscopic  
ensemble of the beam in the ring with $N_{\lambda}'$ the number of particles 
of the spin projection $\lambda$ yielding: 
\begin{equation}
\dot{N}_{-\lambda}' =~\sigma_{-\lambda, \lambda, \lambda^l_i}  
                        j^{l \prime}_{\lambda^l_i} N_{\lambda}', 
\label{eq:thw0}
\end{equation}
where $j^{l \prime}_{\lambda^l_i} = n^{l \prime}_{\lambda^l_i}
\beta'_{\text{relative}} c$ is the current density of the lepton, 
$n^{l \prime}_{\lambda^l_i}$ is the lepton density with the spin projection
${\lambda^l_i}$, and $\beta'_{\text{relative}}c$ is the relative velocity of
the leptons against the hadrons. For easier notation we introduce the
``helicity'' $h_{\lambda^l_i} = n^{l \prime}_{\lambda^l_i}/n^{l \prime}_0$,
the fraction of leptons of the spin projection ${\lambda^l_i}$ of the total
lepton density $n^{l \prime}_0 = n^{l \prime}_+ + n^{l \prime}_-$. 
(The helicity is defined in different ways in the literature, but it is 
convenient to define it here in this way.) We realise that $h_{\lambda^l_i} 
= (1/2)(1 + \lambda^l_i P_l)$ where $P_l$ is the absolute polarisation of 
the macroscopic lepton beam. 

Now we Lorentz transform eq.\,(\ref{eq:thw0}) to the ring frame indicated 
by unprimed quantities with $\beta_{\text{beam}}c$ the velocity of
the hadrons in the rest frame of the ring. Firstly, the lepton density
is given by $n^{l \prime}_0 = n^l_0 \gamma_{\text{beam}}$ due to the length
contraction. Secondly, the time dilatation gives a factor of 
$1/\gamma_{\text{beam}}$ on the right hand side of eq.\,(\ref{eq:thw0}) 
cancelling with the previous one. The transformation of the relative velocity  
$\beta'_{\text{relative}}c$ gives:
\begin{align}
\beta'_{\text{relative}} & = 
\frac{\beta_{\text{relative}}}{1-\beta_{\text{beam}} 
(\beta_{\text{beam}} \pm \beta_{\text{relative}})} \\
& \approx  \frac{\beta_{\text{relative}}}{1-\beta_{\text{beam}}^2} 
=\beta_{\text{relative}} \gamma_{\text{beam}}^2 
\end{align}
if $\beta_{\text{relative}} \ll \beta_{\text{beam}}$. We also observe that
the integrated cross sections are scalars and the same in all reference
systems. Further, $N_{\lambda^h}'=N_{\lambda^h}$ is conserved. 
 
The rate of eq.\,(\ref{eq:thw0}) has to be multiplied by the ratio of 
the interaction length $\ell$ of the lepton beam with the antiproton beam 
over the circumference of the ring $L$. If $f_h$ is the hadron
revolution frequency one has $\ell/L = f_h \ell/(\beta_{\text{beam}} c)$.  
Putting all together eq.\,(\ref{eq:thw0}) reads:
\begin{equation}
\dot{N}_{-\lambda} =~\sigma_{-\lambda,\lambda, \lambda^l_i}\, 
                       n^l_0 h_{\lambda^l_i}\, f_h \,\ell\, 
\frac{\beta_{\text{relative}}}{\beta_{\text{beam}}} 
\gamma^2_{\text{beam}}\, N_{\lambda} \\ 
\label{eq:thw1}
\end{equation}
Since the lepton beam has a finite polarisation we have to add the rates of
the two lepton-spin projections:
\begin{align}
\dot{N}_{-\lambda} &= \sum_{\lambda^l_i} \sigma_{-\lambda,\lambda,\lambda^l_i}\,  
                             h_{\lambda^l_i}\,k_h \, N_{\lambda} \\
                   &= \kappa_{-\lambda}\, N_{\lambda} 
\end{align}
where
\begin{equation}
k_h = n^l_0 f_h \,\ell \,  
\frac{\beta_{\text{relative}}}{\beta_{\text{beam}}} \gamma^2_{\text{beam}}.  
\label{eq:thw2}
\end{equation}

We can now write down the overall accounting of the rate of change of spin 
projections. Since no hadrons are scattered out of the ring $N_+ + N_- =
N_0$ with $N_0$ the total constant number of hadrons coasting in the 
ring we have to observe $\dot{N}_+ = - \dot{N}_-$. This means we have to 
subtract from the rate of change of one spin projection the rate of change 
of the opposite spin projection:
 
\begin{gather}
\dot{N}_{+}= \kappa_+ N_- - \kappa_- N_+ \label{eq:thw3} \\
\dot{N}_{-}= \kappa_- N_+ - \kappa_+ N_- \label{eq:thw4}
\end{gather}
with
\begin{gather}
\kappa_+ = (-2 \Szm P_l - \Sz)\, k_h \\
\kappa_- = (+2 \Szm P_l - \Sz)\, k_h
\end{gather}

It is again convenient to introduce the helicities of the hadron beam $H_+ =
N_+/N_0$ and $H_-=N_-/N_0$ with the beam polarisation $P_h$ yielding:
\begin{equation}
\dot{P_h} = 2 \kappa_+ H_- - 2 \kappa_- H_+
\end{equation}
From this one finds the differential equation for the beam polarisation 
build-up
\begin{equation}
\dot{P_h} = (\kappa_+ - \kappa_-) - (\kappa_+ + \kappa_-) P_h             
\label{eq:P_h_diff_eq}
\end{equation}
with the solution
\begin{multline}
P_h(t) = 2 P_l \frac{\Szm}{\Sz}\{1 - \mathrm{exp}[- \kappa_h t]\}\\ - 
P_h(t=0)\,\mathrm{exp}[- \kappa_h t]
\label{eq:P_h_sol}
\end{multline}
with
\begin{equation}
\kappa_h = 2 |\Sz| k_h.  
\end{equation}
It is evident from these formulae that the maximal polarisation is given by 
$P_h^{\text{max}} = 2 |P_l \frac{\Szm}{\Sz}|$. One reads from 
Fig.\,\ref{ratiospin-flip+} the somewhat disappointing $P_h^{\text{max}}=0.28$
below 100\,keV given by nature. We note that the polarisation build-up is
determined by $\Sz$ only.

The intense beam of an electron cooler is unpolarised. Since the calculation 
of the spin-flip cross section cannot be performed at a kinetic energy 
$T_{h} \rightarrow 0$ as given for the cooler, one may worry about the 
depolarising effect of the cooler. For the antiprotons in focus here, the 
spin-flip cross sections with electrons are many orders of magnitude 
smaller due to the Coulomb repulsion (see section~\ref{sec:crossection}).
Even if we take the maximal cross sections for electrons the achievable
current density in electron coolers of about 10\,mA/mm$^2$ yield 
$\kappa_{\text{cooler}} \approx 1/(10,000$\,h) for a realistic storage 
ring, much too low to produce a noticeable effect.
For the proposed test experiments (see section~\ref{sec:conc}), however, 
this point may have to be considered by switching the cooler on and off.     
However, the classical limit suggests that $\Szm \rightarrow 0$ for $T_{h} 
\rightarrow 0$.

\subsubsection{Polarisation build-up with a pulsed positron source.} 
\label{sec:build-up-pulsed}
With the dc-positron source we can assume that the polarisation of the   
lepton beam is constant since it is refed permanently and polarisation 
loss is negligible by its one path through the interaction region.  
In the situation of two overlapping storage ring beams, the beams interact for
long times. The polarising lepton beam is not anymore a source with constant 
polarisation, but looses polarisation too and regains polarisation from the
hadrons. If left for a sufficiently long time finally an equilibrium
between the two beams will occur. However, the point is to refill the lepton 
ring as often as possible in order to maximise the efficiency. In order to
obtain the coupled differential equation for the lepton polarisation it
suffices to exchange hadrons and leptons in eq.\,(\ref{eq:P_h_diff_eq}) 
since the cross sections are invariant under the exchange 
$\lambda^h_f \rightarrow \lambda^l_f,~\lambda^h_i \rightarrow \lambda^l_i$ 
and $\lambda^l_i \rightarrow \lambda^h_i$. 
We define: 
\begin{equation}
k_l = n^h_0 f_l \,\ell \, 
\frac{\beta_{\text{relative}}}{\beta_{\text{beam}}} \gamma^2_{\text{beam}}
\end{equation}
and get:
\begin{alignat}{4}
\dot{P_h} &=& -4 k_h \Szm &P_l  + 2 k_h \Sz &P_h &= \mu_h &P_l - \nu_h &P_h
\label{eq:Ph_pulsed}\\
\dot{P_l} &=& -4 k_l \Szm &P_h  + 2 k_l \Sz &P_l &= \mu_l &P_h - \nu_l &P_l
\label{eq:Pl_pulsed}
\end{alignat}
The solution of this coupled equations is trivial, but the expressions are
lengthy and do not provide much insight. This is inconvenient since the
solutions of eqs.\,(\ref{eq:Ph_pulsed}) and (\ref{eq:Pl_pulsed}) have to be 
applied iteratively following the numbers of refills of the lepton ring. 
This means inserting at the beginning of each refill the new initial hadron 
polarisation and the new initial lepton polarisation and then solve the 
equations anew. We shall therefore show solutions for the numerical 
parameters for the design example presented in 
section~\ref{sec:positron-source-pulsed} in graphical form.

\subsection{Design examples}       
\label{sec:app}
The different kinds of polarised positron sources offer several applications 
of polarising antiprotons in storage rings. In this subsection we give two 
examples of an application of the parallel beam method. The first is a 
specialised polariser ring together with the dc-source sketched in 
subsection~\ref{sec:positron-source-dc}. It is optimised for fast polarisation  
build-up, adequate for an external fixed target experiment requiring a 
slowly extracted beam. It would be suit for the experiments mentioned in the 
introduction under ``1. Spectroscopy of hadrons'' and ``2. Antinucleon-nucleon
scattering and reactions'' in the introduction. The second is the idea to use 
a small positron storage ring the beam of which overlaps with a storage ring 
similar to the HESR in ref.~\cite{Len05}. This configuration was already 
discussed in subsections \ref{sec:positron-source-pulsed} and 
\ref{sec:build-up-pulsed}. It would be suitable for ``3. Antinucleon-nucleon 
interactions at the parton level'' in order to allow for a direct comparison 
of the performance  with the scheme presented in ref.~\cite{Rat05}. It is, 
however, not the purpose of this paper to present elaborated proposals. 
There are too many interlinked parameters and constraints coming from the 
specific experiments and the accelerator limitations. Some of the limitations 
and needs for studies in the future will be discussed in 
section~\ref{sec:conc}.  

\begin{figure}[ht]
\centering{
\includegraphics[width=0.8\columnwidth]{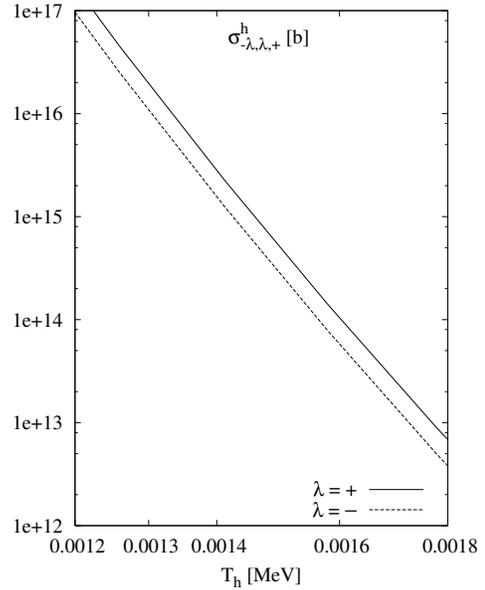}}
\caption{The integrated spin-flip cross sections 
  $\langle \sigma^h_{-\lambda,\lambda,+}\rangle$ ($\lambda=\pm$) for proton
  electron scattering in the c.m.\ frame in the region of very low proton 
  lab kinetic energy (magnification from Fig.~\ref{sigspin-flip+}).}
\label{sigspin-flip+_ausschnitt} 
\end{figure}

As discussed in the preceding section~\ref{sec:build-up} the decisive cross
section for the time needed for polarisation build-up is $2\Sz$. Since we 
optimise for short times we chose small relative energies between the lepton 
and hadron. Figure~\ref{sigspin-flip+_ausschnitt} shows the magnification of
Fig.~\ref{sigspin-flip+} suggesting a choice of $T_h = 0.0017\,$MeV yielding
$\Sz = - 2 \cdot 10^{13}$\,barn. This choice guarantees the validity of the DW
calculation (see refs.~\cite{Are07,Are07a}) with a large cross section.

\subsubsection{Specialised polariser ring}                 
\label{sec:specring}
\begin{figure}[h]
\includegraphics[width=0.99\columnwidth]{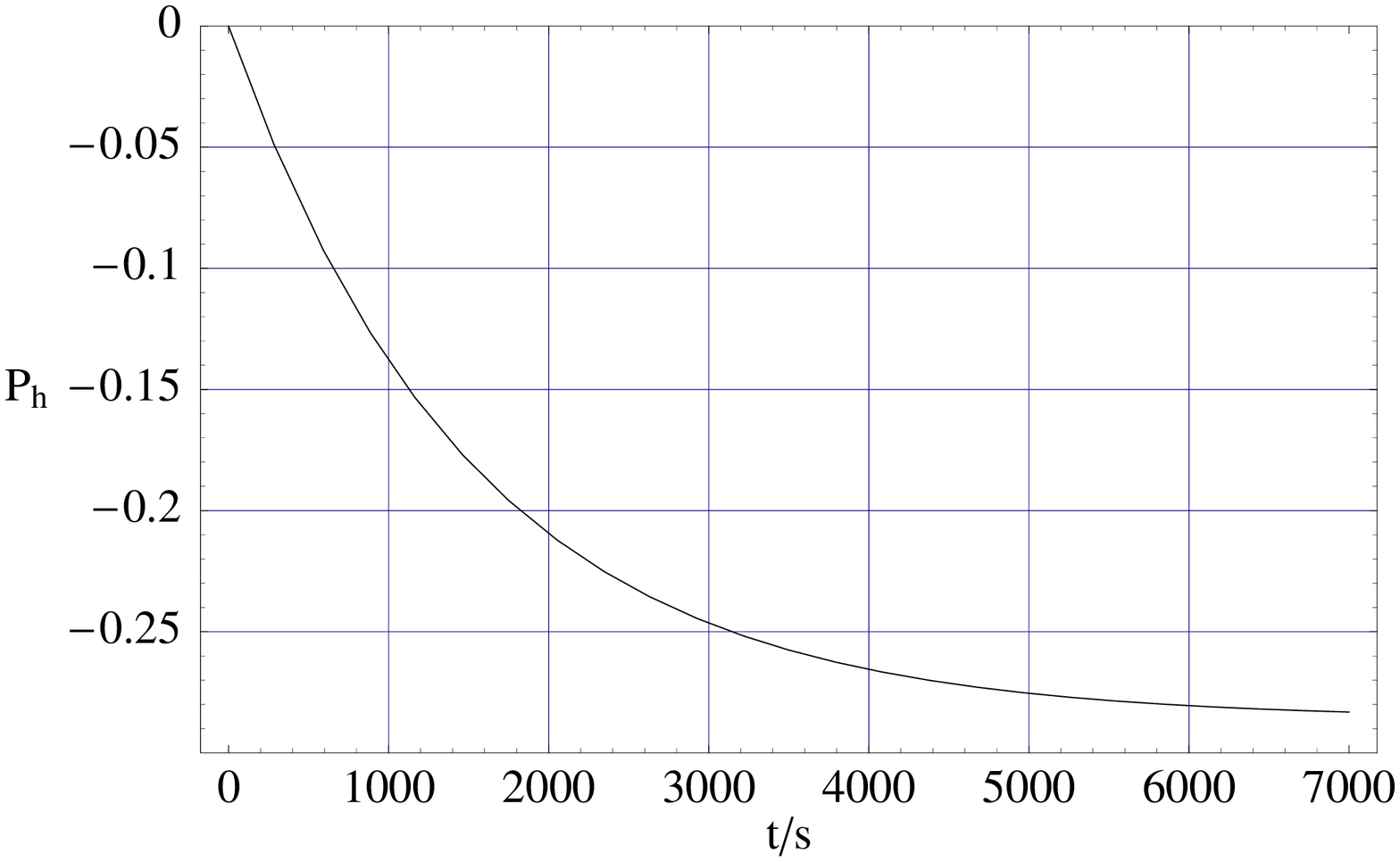}
\caption{The antiproton polarisation build-up for a polariser with $P_l = 1.$ 
  and an initially unpolarised hadron beam $P_h(t=0)=0$. The polarisation 
  has to be multiplied by the positron polarisation of 0.7 for the dc source.} 
\label{fig:P_h} 
\end{figure}

First we want to present a specialised polariser ring optimised for fast
polarisation build-up. We start with the parameters of the example PB$_1$ in 
Table~\ref{tab:table_PB} which is close to cooler rings built before, as e.g.
LEAR (see \cite{Poth:1990pg} and references therein) or the TSR \cite{Rat93}.
This design represents a reasonable compromise between size and cost.
\newcommand{\mc}[1]{\multicolumn{2}{|c|}{#1}}
\begin{table}
\begin{center}
\begin{tabular}{|l||c|c|}
\hline \hline
example  &  PB$_1$ &  PB$_2$ \\
\hline \hline
\multicolumn{3}{|c|}{antiproton beam in ring system} \\
\hline \hline
$\beta_{\text{antiproton beam}}=v/c$ & 0.50 & 0.95\\
\hline
kinetic energy & 145.151\,MeV & 2066.6\,MeV\\
\hline
momentum & 541.712\,MeV/c & 2854.6\,MeV/c \\ 
\hline
rigidity & 1.81\,T\,m & 9.54\,T\,m \\
\hline
circumference L & 75\,m & 400\,m \\
\hline 
revolution frequency & 2\,MHz & 0.71\,MHz \\
\hline
acceptance & \mc{25~$\pi$\,mm\,mrad} \\
\hline
$\beta_0$ at mid overlap & \mc{2.2\,m}   \\
\hline
beam emittance $\pi \varepsilon$ & \mc{0.45\,$\pi$~mm\,mrad} \\
\hline
beam particles & \mc{$10^{10}$} \\
\hline
$\Delta Q_{sp}$ & 0.014 & 0.0002 \\
\hline \hline
\multicolumn{3}{|c|}{positron beam in ring system} \\
\hline \hline
$\beta_{\text{electron beam}}=v/c$ & 0.501426 & 0.950185\\
\hline
kinetic energy & 79.6145\,keV & 1128.47\,keV\\
\hline 
momentum & 295.025\,keV/c & 1557.80\,keV/c\\
\hline
average current $\langle I_e \rangle$ & $5 \cdot 10^{9}e^+s^{-1}$
                                      & $1.2 \cdot 10^{13}e^+ s^{-1}$ \\
\hline
peak current $I_e^{\text{peak}}$ & --  & $6 \cdot 10^5 e^+ (4\,\mu s)^{-1}$  \\ 
\hline
repetition frequency & dc & 28\,Hz \\
\hline
positron density $n_e$ & $1.1 \cdot 10^{7}$\,m$^{-3}$ & 
                         $1.4\cdot 10^{10}$\,m$^{-3}$ \\
\hline
trans. emittance $\pi \varepsilon_e$ & 0.45\,$\pi$\,mm\,mrad & 3\,$\pi$\,mm\,mrad\\
\hline
beam diameter & \mc{2\,mm} \\
\hline
overlap length $\ell$ & \mc{2\,m} \\
\hline \hline
\multicolumn{3}{|c|}{relative motion of electron in antiproton frame}\\
\hline \hline
$\beta'_{\text{relative}}$ & \mc{0.0019036} \\
\hline
kinetic energy $T_p$ & \mc{1.7\,keV} \\
\hline
kinetic energy $T_e$ & \mc{0.93\,eV} \\
\hline
$\Sz$ & \mc{$-2 \cdot 10^{13}$\,barn} \\
\hline
$\Szm/Sz$ & \mc{-0.143} \\
\hline 
$\tau_{\text{pol.}} = 1/\kappa_h$ & 0.43~h & 6.4~s \\
\hline
$\tau_{\text{pol. build-up}}= \tau_{\text{pbu}}$ & 1~h & 1~h \\
\hline
polarisation $P_l$ & 0.70 & 0.76 \\
\hline
polarisation $P_h(\tau_{\text{pbu}})$ & 0.18 & 0.17  \\
\hline \hline 
\end{tabular}
\caption{Two design examples of polarising antiprotons with the parallel beam  
         method (PB) in a storage ring using discussed parameters for the
         positron sources. $T_p$ and $T_e$ are the kinetic energies of the
         antiproton and positron in the rest frame of the respective other
         particle.}  
\label{tab:table_PB} 
\end{center} 
\end{table}
At the relative kinetic energy of $T_p = 1.7$\,keV we have the ratio
$\Szm/\Sz=-0.143$ (see Fig.~\ref{ratiospin-flip+}). If we assume a freely 
coasting antiproton beam and, in accord with the discussion in 
subsection~\ref{sec:positron-source-dc}, an average positron current of
$\langle I_e \rangle = 5 \cdot 10^{9}e^+s^{-1}$ and all other parameters 
as given in Table~\ref{tab:table_PB}, we get from eq.\,(\ref{eq:thw2}) 
$\kappa_h$ = 2.38\,h$^{-1}$. Assuming a positron polarisation of 0.70 we 
get according to eq.\,(\ref{eq:P_h_sol}) (see Fig.~\ref{fig:P_h}) 
an antiproton polarisation of 0.18 after one hour. 

The coasting beam in a storage ring with the parameters assumed for 
PB$_1$ in Table~\ref{tab:table_PB} represents no problem for the particle 
dynamics. A few aspects are however mentioned in the following.
The incoherent tune shift spread $\Delta Q$ for our design example is 
calculated with eq.\,(\ref{spch}). For $N^{\Delta Q} = 10^{10}$ we 
get $\Delta Q = 0.013$ well in accord with running conditions in existing 
similar storage rings. $\Delta Q$ is one of the critical parameters 
of the parallel beam method and limits the number of antiprotons in the ring. 
It may be possible to go to larger $\Delta Q$ than 0.013 in a specially 
optimised ring lattice but we did not investigate such an optimisation.   
 
This means that the polariser ring could be filled at the rate of 
$10^7 \bar{p}$/s in about 16 minutes with $10^{10}$ antiprotons and then 
polarised in 1 hour, i.e. two polarisation times 
$\tau_{\text{pol}} = 1/\kappa_h$, to a polarisation of 0.18.

The relative kinetic energy $1.7$\,keV has been chosen to be above the limit of
validity of the DWBA calculations. Since the cross sections depend smoothly on
the energy no narrow limits are set for the beam spread.  Requesting a 
relative momentum spread of $\delta p/p = \pm 1\cdot 10^{-4}$ in the ring
system for both the antiproton as well as for the positron, standard in modern  
storage rings, one arrives after Lorentz transforming at an energy spread 
of $\delta T_{\bar{p}} \lessapprox |\pm 0.1| T_{\bar{p}} = |\pm 0.17|$\,keV 
in the rest system of the electron. These numbers for the following example 
PB$_2$ are so similar that they are not discussed separately. 

The region of overlap of the antiproton beam and polarised positron beam is
assumed to have a length of 2\,m. In order to achieve a good efficiency we
need a complete overlap with the smallest diameter possible. The diameter 
we have assumed is 2\,mm. As discussed for the dc-source this poses no 
problem. However, if we want to maintain a diameter for the pulsed source 
we have to provide a focusing by means of a solenoid. The space charge 
effects are small and a guiding field of less than 0.01\,T suffices to 
maintain the beam diameter of 2\,mm. 

For this beam diameter we estimate a minimal value of the beta function 
$\beta_{min}$ of the antiproton beam in mid overlap according to 
\begin{equation}
\beta_{min} = \frac{l/2}{\sqrt{p^2-1}},
\end{equation}
where $ p = d_e/d_0 $ with $d_e$ the beam diameter at the entrance and exit 
and $d_0$ the diameter at the mid point of the overlap region. With $p = 1.1$ 
of we obtain $\beta_0 >$2.2\,m. With this value and a beam diameter of 2\,mm 
we get from the relation 
\begin{equation} 
\sigma_{x,y} = \sqrt{\beta_{x,y}\,\varepsilon_{x,y}}
\end{equation}
$\varepsilon_{x,y}$ = 0.45\,mm\,mrad as the needed emittance for the 
antiproton beam. This means that we have to cool the antiproton beam after 
injection before we can efficiently polarise. The cooling has to stay on
during polarisation build-up in order to compensate for the intra-beam
scattering and the multiple scattering in the rest gas of the ring.
Emittances of about 0.1\,mm\,mrad have been obtained at LEAR and other cooler
rings for up to $10^{9}$ particles in the ring \cite{Poth:1990pg}. 

\subsubsection{HESR as polariser ring}                         
\label{sec:HESR}
The principle of the dc-source could be used by adjusting the parameters 
of the previous design example PB$_1$ to a high energy ring as HESR.
On the other hand, the pulsed source offers, as repeatedly mentioned, the 
interesting option of letting the beam of a positron storage ring overlap 
with the antiproton beam. This will be investigated in the following.   

\begin{figure}[h]
\begin{minipage}[t]{0.98\columnwidth}
 \includegraphics[width=0.98\columnwidth]{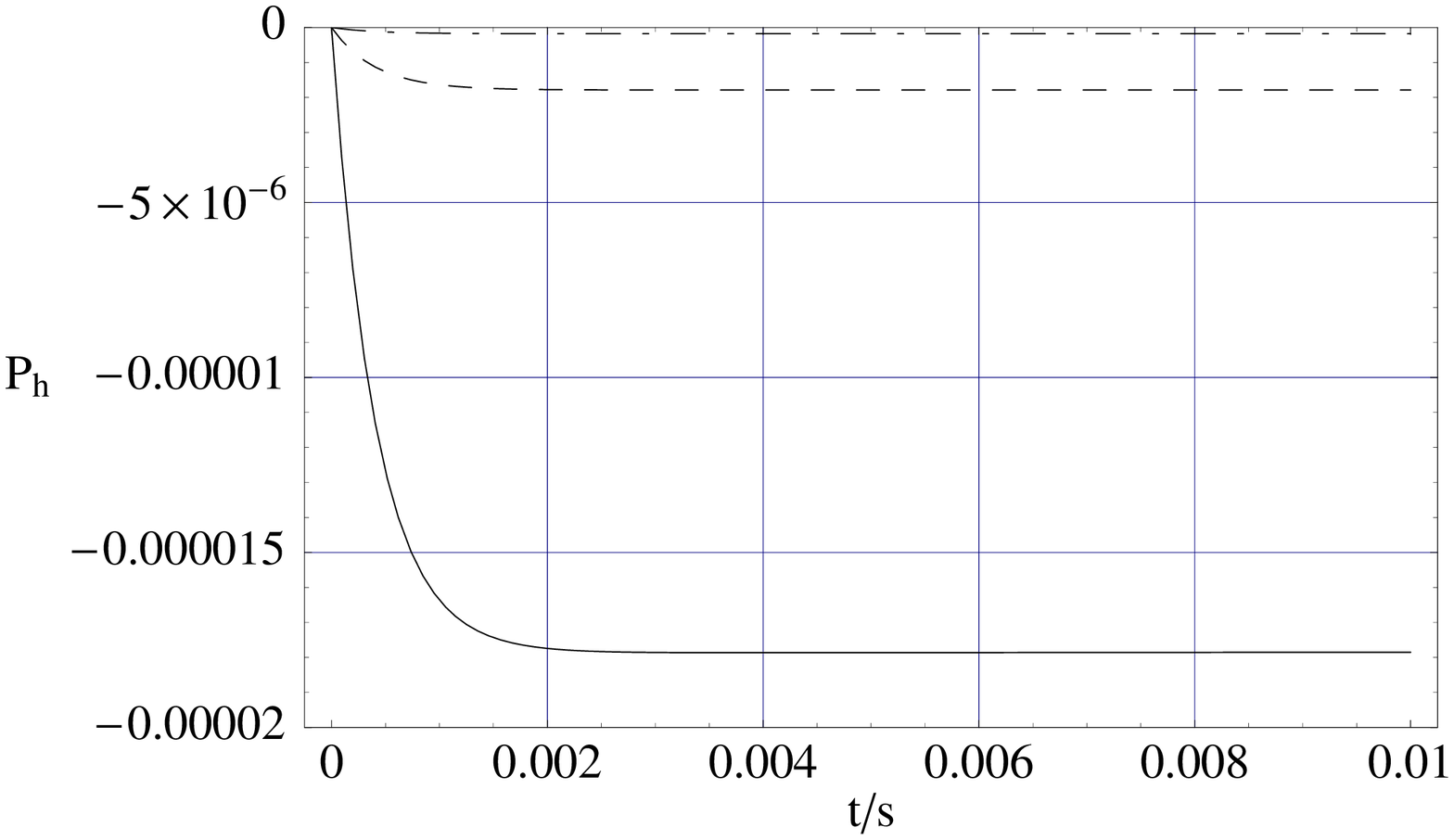}
\caption{The polarisation build-up $P_h$ for a single fill with the starting
  hadron polarisation $P_h = 0$ and lepton polarisation $P_l=1$. The full
  curve is for the parameters in Table\,\ref{tab:table_PB}, i.e. 
  $a=N^h_0/N^l_0=1.6 \cdot 10^4$, the dashed curve for $a=1.6 \cdot 10^5$, and 
  the dashed dotted for $a=1.6 \cdot 10^6$.}
\label{fig:Ph_pulsed_0}

\vspace{4mm}
 \includegraphics[width=0.98\columnwidth]{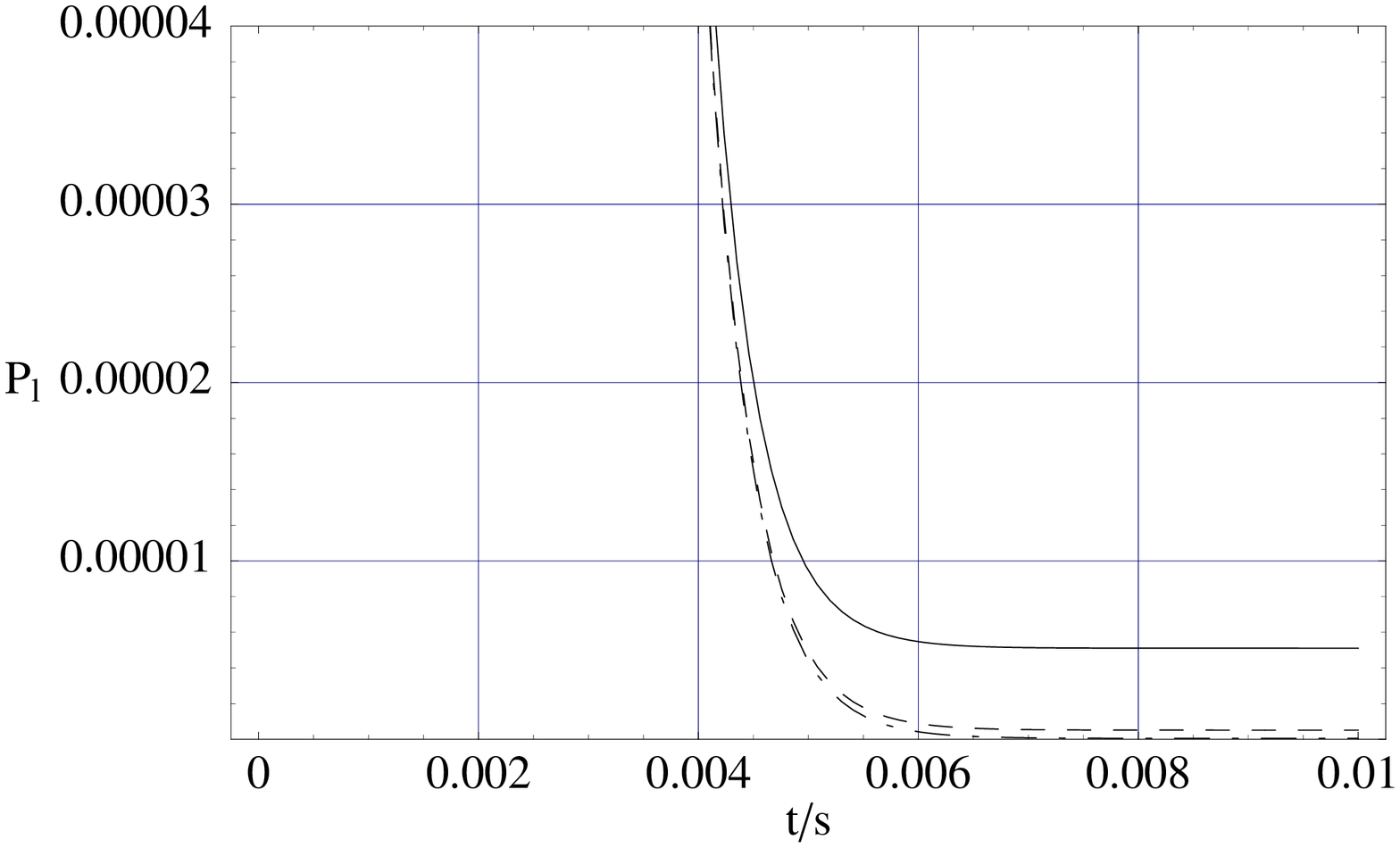}
\caption{The polarisation build-up $P_l$ for a single fill with the starting
  hadron polarisation $P_h = 0$ and lepton polarisation $P_l=1$ . 
  Curves as in fig.\,\ref{fig:Ph_pulsed_0}.}
\label{fig:Pl_pulsed_0} 
\end{minipage}
\end{figure}

\begin{figure}[h]
\begin{minipage}[t]{0.98\columnwidth}
 \includegraphics[width=0.98\columnwidth]{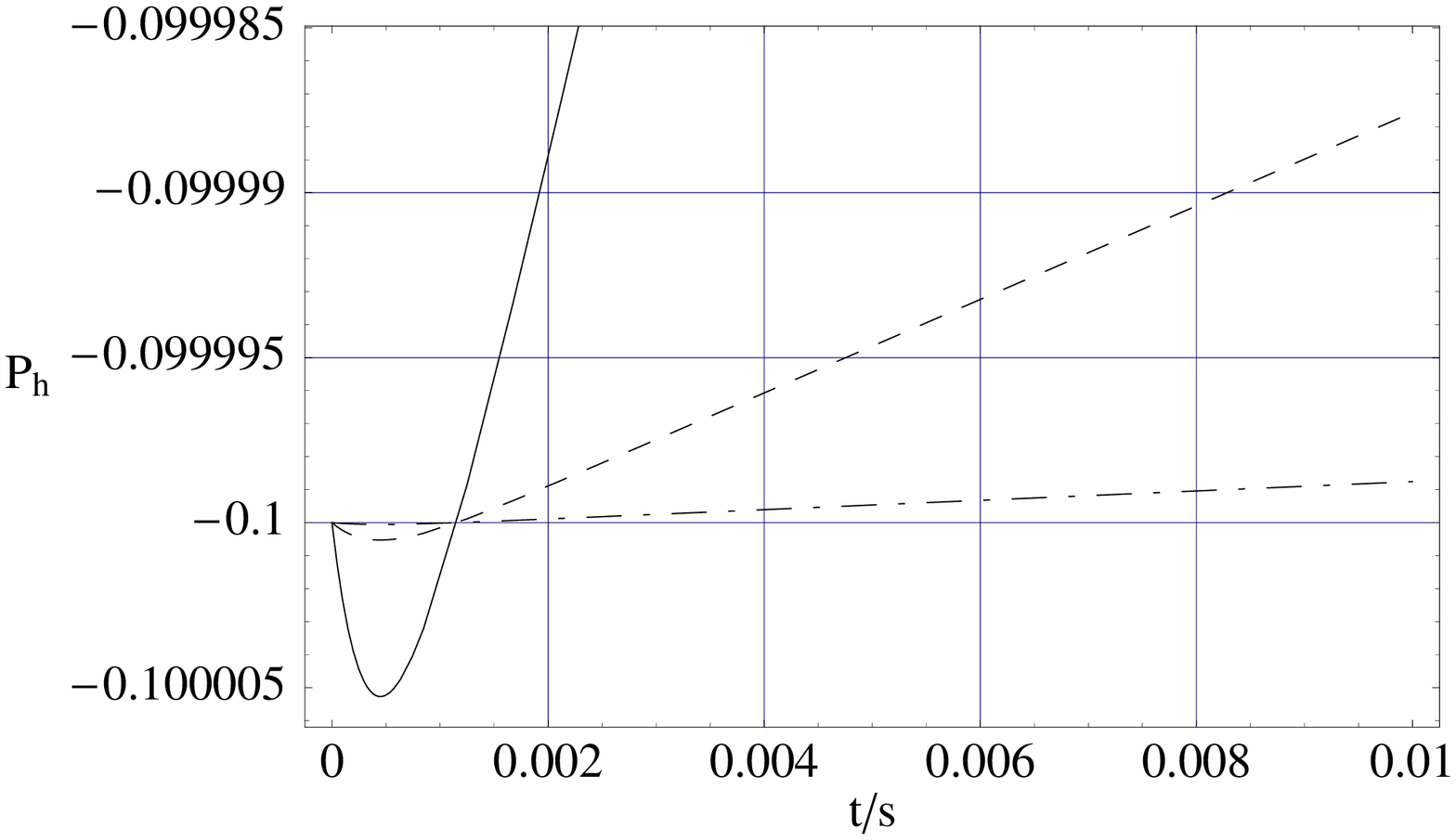}
\caption{The polarisation build-up $P_h$ for a single fill with the starting
  hadron polarisation $P_h = -0.1$ and lepton polarisation $P_l=1$. 
  Curves as in fig.\,\ref{fig:Ph_pulsed_0}.}
\label{fig:Ph_pulsed_0.1}

\vspace{4mm}
\includegraphics[width=0.99\columnwidth]{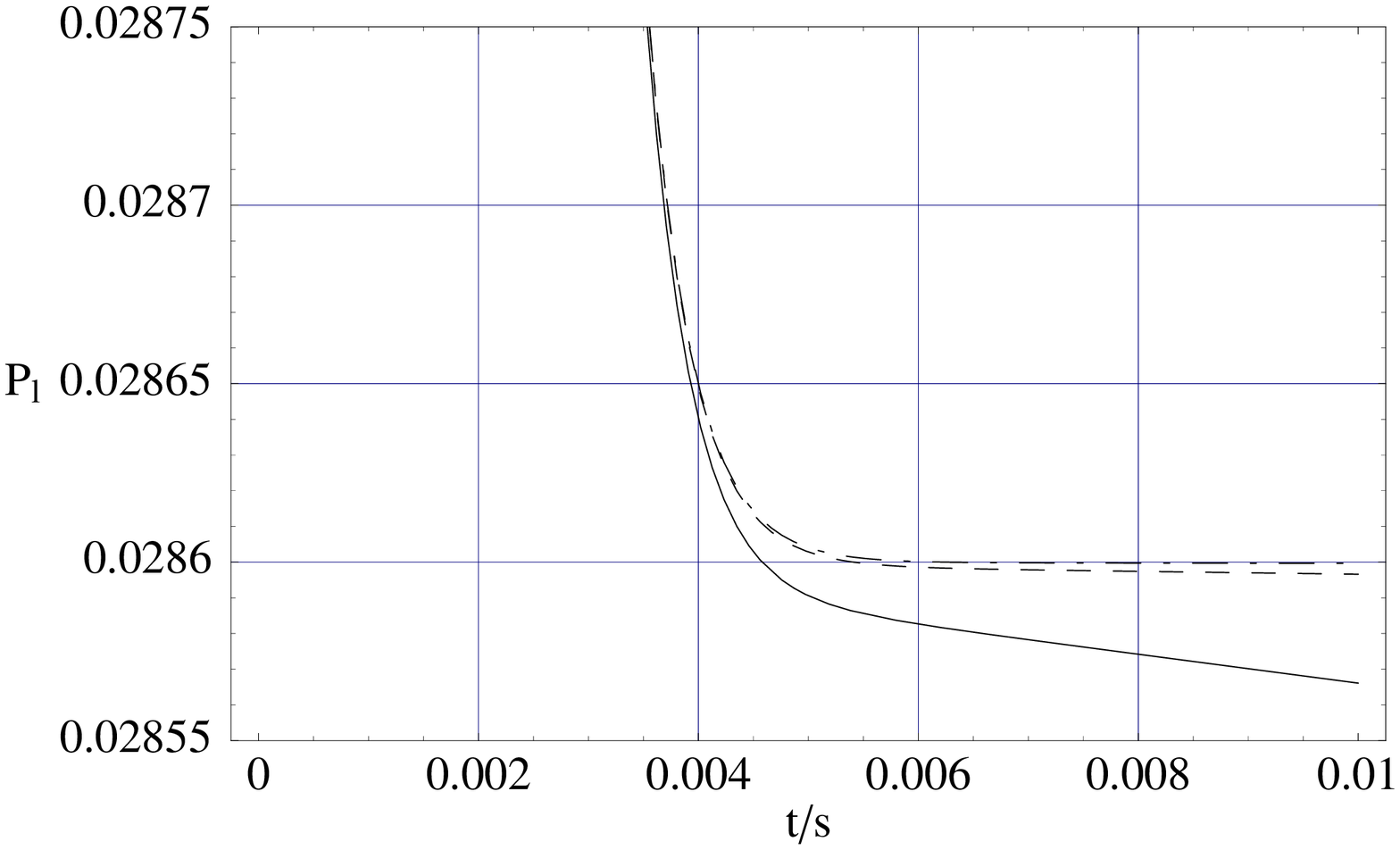}
\caption{The polarisation build-up $P_l$ for a single fill with the starting
  hadron polarisation $P_h = -0.1$ and lepton polarisation $P_l=1$. 
  Curves as in fig.\,\ref{fig:Ph_pulsed_0}.}
\label{fig:Pl_pulsed_0.1} 
\end{minipage}
\end{figure}

\begin{figure}[h]
\includegraphics[width=0.99\columnwidth]{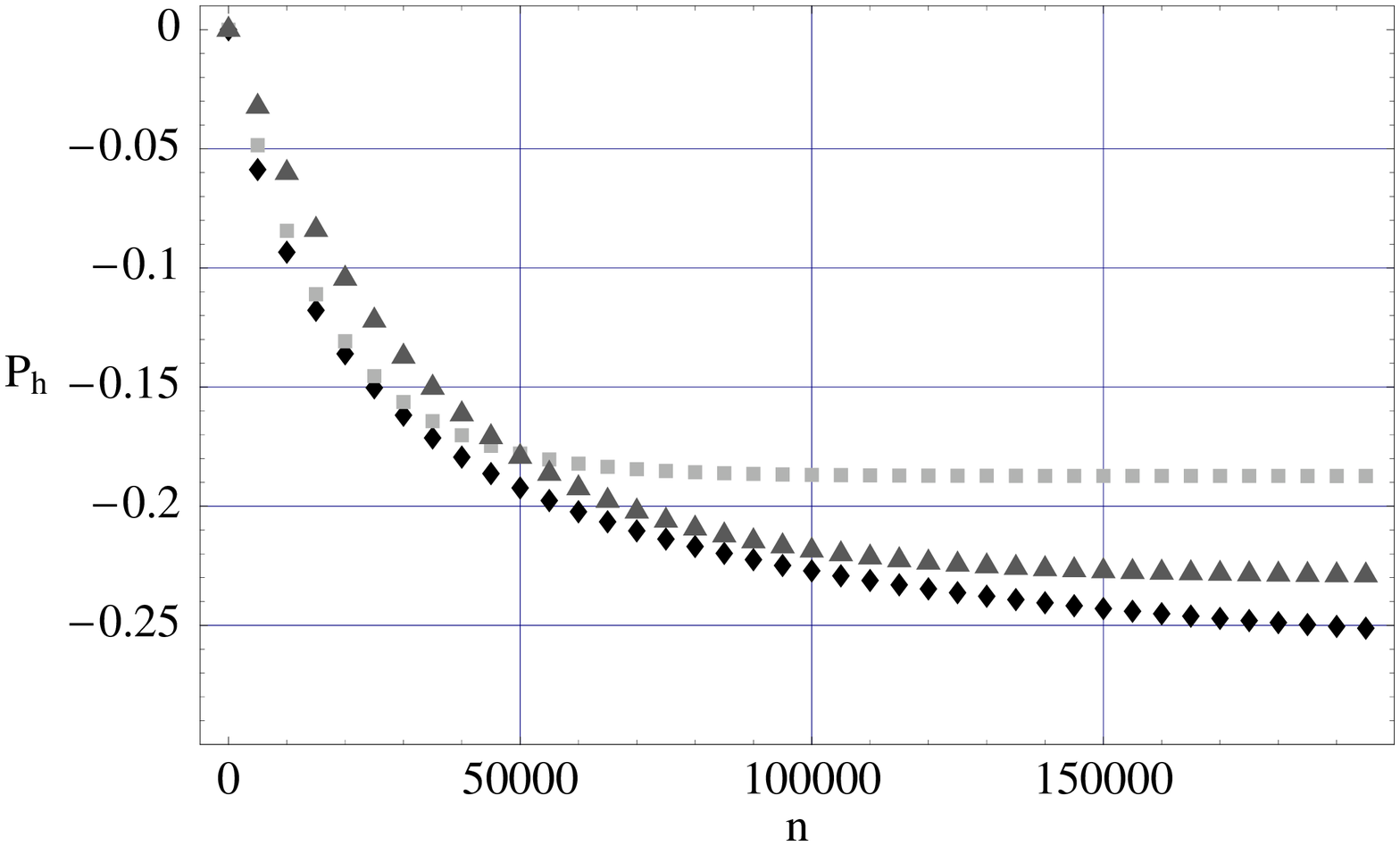}
\caption{The polarisation build-up $P_h$ as a function of the number of lepton
  refills (cycles) for $P_h(t=0) = 0$, $P_l(n \cdot \delta \tau) = 1$, and 
  $a = N^h_0/N^l_0 = 1.6 \cdot 10^4$. The polarisation time for one cycle 
  $\delta \tau$ is 0.4\,ms (light grey boxes) and 0.2\,ms (dark grey
  triangles). The black diamonds show a solution for an optimised polarisation
  time depending on the number of cycles ranging from $\delta \tau =
  1.7$\,ms at $P_h(1)=0$ for the beginning of the first cycle to 
  $\delta \tau = 0.16$\,ms at $P_h(6 \cdot 10^4)=-0.2$ for the
  beginning of the $6 \cdot 10^4$th cycle. The cycles repeat with a frequency 
  of 28\,Hz in order to limit the average current of the source of polarised 
  electrons.}  
\label{fig:Ph_pulsed_total} 
\end{figure}

We start by observing that
\begin{gather}
n^l_0 = N^l_0/(A\,L_l)\\
n^h_0 = N^h_0/(A\,L_h)
\end{gather}
and define:
\begin{equation}
a = \frac{k_l}{k_h} = \frac{(N^h_0 / L_h) f_l/ \beta_h}{(N^l_0/L_l) f_h/ \beta_l}
= \frac{N^h_0}{N^l_0}
\end{equation}
For the parameters collected in Table\,\ref{tab:table_PB} for the design
example PB$_2$ $\nu_h = 2 k_h |\Sz| \approx 2/(6.4$\,s) and  
$a = 1.6 \cdot 10^4$ giving $\nu_l = a \nu_h \approx  1/(0.38$\,ms). The
ratio $\mu/\nu= 2 \Szm/\Sz$ can be determined with Fig.~\ref{ratiospin-flip+}. 
We plot the solutions of eqs.\,(\ref{eq:Ph_pulsed}) and (\ref{eq:Pl_pulsed}) 
with these parameters in Figs.\,(\ref{fig:Ph_pulsed_0}), (\ref{fig:Pl_pulsed_0}),
(\ref{fig:Ph_pulsed_0.1}), and  (\ref{fig:Pl_pulsed_0.1}).
We can see that the polarisation build-up gets increasingly inefficient with 
growing initial polarisation. This is due to the exchange of spin and angular 
momentum between the hadron and lepton beam. The total spin in the lepton beam 
is not preserved since the tensor part of the hyperfine interaction transfers 
spin into angular momentum. For large times both polarisations converge to
zero. We also observe that the optimum time, after which the maximum
polarisation increase is reached, changes with the number of cycles, i.e. 
initial polarisation of the hadron beam. It is efficient to stop the
polarisation cycle at this optimum time and wait until the lepton source can
deliver the next spill for injection into the lepton ring. We show the
iterative polarisation build-up as a function of the number of cycles in 
Fig.\,\ref{fig:Ph_pulsed_total}. If one changes the optimum time from lepton 
spill to lepton spill one gains somewhat as also shown in 
Fig.~\ref{fig:Ph_pulsed_total}.  

If we chose $\delta \tau = 0.2\,ms$ with a repetition (cycle) time 
of $\tau_{cyle} = 36$\,ms we get after $10^5$ cycles a total time for
polarisation build up of one hour with a polarisation of $P_h=0.22 \cdot
P_l(t=0)$. Considering the polarisation of the positrons $P_l = 0.76$ we get 
an effective polarisation of the antiproton beam of 0.17.

It is worth noting that one can use the scheme of this section with a cross
section $\Sz$ of up to a factor of hundred smaller than predicted by the DW
calculation. The time dependence in Figs.~\ref{fig:Ph_pulsed_0},
\ref{fig:Pl_pulsed_0}, \ref{fig:Ph_pulsed_0.1}, and \ref{fig:Pl_pulsed_0.1} 
just scales with $1/Sz$ and the polarisation build-up during one cycle is 
correspondingly slower. This means that also $\delta \tau \propto 1/\Sz$ and 
one can go up to $\delta \tau \approx \tau_{cycle}$. We remind that 
$\tau_{cycle}$ is limited by the maximal possible average current of the 
source of polarised electrons.

\section{Discussion and Conclusions}                            
\label{sec:conc}
Now we want to compare the internal target method (IT) \cite{Rat05} with the
parallel beam method (PB) of this article on the basis of the figure-of-merit
FOM\,=\,$N_{0\,b} \cdot P_b^2/\tau_{\text{pbu}}$, where we mean with 
$\tau_{\text{pbu}}$ the time for one cycle of polarising the antiprotons
and using them in an experiment. If the time needed to perform the
experiment is shorter or comparable to the polarisation cycle time a 
reduction of the overall efficiency is indicated which is, however,
difficult to reflect in a FOM since it depends on the specific 
experiment. 

For the internal target method we assume a filling time of one hour since it 
needs many antiprotons in order to make up for the many particles lost due to 
Coulomb scattering and annihilation. This means that 9 hours are left for 
polarisation build-up. Further, the numbers of Table~\ref{tab:Rat05} are used. 

For the parallel beam method PB$_1$, i.e. polarisation in a separate polariser
ring, we can fill the ring in 16.7 minutes and consequently fill and polarise
in one hour. With a polarisation time of $\tau_{\text{pol.}}=1$\,hour we reach 
a polarisation of the antiprotons of $P_h = 0.18$.   

For the parallel beam method PB$_2$ we take as polarisation build-up time 
$\tau_{\text{pbu}}=1$\,hour using the HESR as polarisation ring. This means 
that one cannot use the ring for experiments about half of the time assuming 
a beam life time of two hours in HESR. However, if one would polarise and
perform the experiment in parallel, one could do routine detector checks
mandatory for precision measurements in the first half of the life time of
small polarisation and do the real experiment in the second half. On the 
other hand one can easily change the spin direction for the antiprotons with 
each filling, a feature highly desirable.

Table~\ref{tab:fom} shows the comparison of all examples.
\setlength{\tabcolsep}{5pt}
\begin{table}[h]
\centering
\begin{tabular}{|l||c|c|c|c|c|}
\hline
method & IT$_{a}$ & IT$_{b}$ & IT$_{c}$ & PB$_1$ & PB$_2$ \\ 
\hline \hline
$N_{pol}$ & $2 \cdot 10^{10}$ & $5 \cdot 10^9$ & $2 \cdot 10^7$
& $1 \cdot 10^{10}$ & $1 \cdot 10^{10}$ \\ 
\hline
$P_h$  & 0.11 & 0.29 & 0.46 & 0.18 & 0.17 \\
\hline
$\tau_{\text{pbu}}$ /h & 10 & 10 & 10 & 1 & 1 \\
\hline
FOM & $2.6 \cdot 10^7$ & $4.2 \cdot 10^7$ & $4.2 \cdot 10^5$ & $3.2 \cdot 10^8$
& $2.9 \cdot 10^8$ \\
\hline
\end{tabular}
\caption{Figure-of-merit FOM\,=\,$N_b \cdot P_b^2/\tau_{\text{pbu}}$ 
  for the internal target method (IT) with parameters as in 
  Table~\ref{tab:Rat05} and for the parallel beam method (PB) with 
  parameters as in Table~\ref{tab:table_PB} (see also text). The polarisation 
  for the internal target method has been recalculated using eq.\,(4) of
  ref.~\cite{Rat05}. $\tau_{\text{pbu}}$ is the time needed for one polarisation
  cycle in an experiment.}
\label{tab:fom}
\end{table}
The parallel beam method for polarising antiprotons has figure-\-of-\-merits 
about one order of magnitude better than the internal target method. 
Unfortunately, the natural limit given by the theoretically predicted 
spin-flip cross sections $\Sz$ and $\Szm$ does not allow to surpass a 
maximum polarisation of 0.28. Nevertheless, it appears that the parallel 
beam method is superior over any other method proposed so far, since it 
provides a still reasonable polarisation and a large figure-of-merit at 
low investments. It should be experimentally verified at an existing 
storage ring as soon as possible in view of the importance for the 
planning at FAIR. Fortunately the electron-proton interaction has 
unlike signs as the one of positron-antiproton. If one would inject 
polarised protons in a storage ring and detune the frequently available 
electron cooler somewhat one could measure the depolarisation time and 
determine the spin-flip cross section $\Sz$ (see eq.\,(\ref{eq:P_h_sol}) 
with $P_l=0$ and $P_h(t=0) \neq 0$). However, the spin-flip cross section
$\Szm$ can only be determined using polarised leptons. Such a test experiment 
would completely exclude any risk of an expensive design solely based on 
the predictions.

\begin{acknowledgement}
This work has been supported by the SFB~443 of the Deutsche
For\-schungs\-ge\-meinschaft (DFG) and the German Federal State of
Rhein\-land-Pfalz. 
\end{acknowledgement}

\def\etal{\textit{et al.}}
\def\journ#1#2#3#4{{#1}\textbf{#2}, #3 (#4)}
\def\EPJA#1#2#3{\journ{Eur.\ Phys.\ J.\ A\ }{#1}{#2}{#3}}
\def\PRL#1#2#3{\journ{Phys.\ Rev.\ Lett.\ }{#1}{#2}{#3}}
\def\PR#1#2#3{\journ{Phys.\ Rev.\ }{#1}{#2}{#3}}
\def\PRA#1#2#3{\journ{Phys.\ Rev.\ A\ }{#1}{#2}{#3}}
\def\PRB#1#2#3{\journ{Phys.\ Rev.\ B\ }{#1}{#2}{#3}}
\def\PRC#1#2#3{\journ{Phys.\ Rev.\ C\ }{#1}{#2}{#3}}
\def\PRE#1#2#3{\journ{Phys.\ Rev.\ E\ }{#1}{#2}{#3}}
\def\PL#1#2#3{\journ{Phys.\ Lett.\ }{#1}{#2}{#3}}
\def\NP#1#2#3{\journ{Nucl.\ Phys.\ }{#1}{#2}{#3}}
\def\NIMA#1#2#3{\journ{Nucl.\ Instrum.\ Meth.\ A\ }{#1}{#2}{#3}}
\def\NIMB#1#2#3{\journ{Nucl.\ Instrum.\ Meth.\ B\ }{#1}{#2}{#3}}
\def\RMP#1#2#3{\journ{Rev.\ Mod.\ Phys.\ }{#1}{#2}{#3}}

\end{document}